\documentclass[%
 reprint,
 amsmath,amssymb,
 aps,
prb,
]{revtex4-1}

\usepackage{graphicx}
\usepackage{dcolumn}
\usepackage{bm}

\usepackage[caption=true]{subfig}
\usepackage{float}

\newcommand{\labeq}[1]{\begin{equation} #1 \end{equation}}
\newcommand{\labalign}[1]{\begin{align} #1 \end{align}}

\newcommand{ \mb }{ \mathbf } 
\newcommand{ \qq }{ \mb{q}} 
\newcommand{ \kk }{ \mb{k} } 
\newcommand{ \GG }{ \mb{\mathcal{G}} } 

\newcommand{ \lam }{ \lambda } 
\newcommand{ \ddt }{ \partial_{t} } 
\newcommand{ \bracom }[2]{ \left( #1 \right)_{\text{#2}} } 
\newcommand{ \fdsub }[1]{ f^{0}_{#1} }
\newcommand{ \besub }[1]{ n^{0}_{#1} }
\newcommand{ \velsub }[1]{ \mb{v}_{#1} }
\newcommand{ \ensub }[1]{ \epsilon_{#1} }
\newcommand{ \gsq }{ \left | g^{smn}_{\kk\qq} \right |^{2}  }
\newcommand{ \elen }[1]{ \epsilon_{#1}  }
\newcommand{ \phen }[1]{ \hbar\omega_{#1}  }
\newcommand{ \Isub }[1]{ \mb{I}_{#1} }
\newcommand{ \Fsub }[1]{ \mb{F}_{#1} }
\newcommand{ \Jsub }[1]{ \mb{J}_{#1} }
\newcommand{ \Gsub }[1]{ \mb{G}_{#1} }
\newcommand{\stirling}[2]{\genfrac{\{}{\}}{0pt}{0}{#1}{#2}}

\newcommand{\ket}[1]{| #1 \rangle}
\newcommand{\bra}[1]{\langle #1 |}

\begin{document}

\title{Coupled transport of phonons and carriers in semiconductors: a case study of n-doped GaAs}

\author{Nakib H Protik}
 \email{nakib@seas.harvard.edu}
 \affiliation{John A. Paulson School of Engineering and Applied Sciences, Harvard University, Cambridge, Massachusetts 02138, USA}
\author{David A Broido}%
 \email{broido@bc.edu}
\affiliation{%
 Department of Physics, Boston College,Chestnut Hill, Massachusetts 02467, USA\\
}%

\date{\today}

\begin{abstract}
We present a general coupled electron-phonon Boltzmann transport equations (BTEs) scheme designed to capture the mutual drag of the two interacting systems. By combining density functional theory based first principles calculations of anharmonic phonon-phonon interactions with physical models of electron-phonon interactions, we apply our implementation of the coupled BTEs to calculate the thermal conductivity, mobility, Seebeck and Peltier coefficients of n-doped gallium arsenide. The measured low temperature enhancement in the Seebeck coefficient is captured using the solution of the fully coupled electron-phonon BTEs, while the uncoupled electron BTE fails to do so. This work gives insights into the fundamental nature of charge and heat transport in semiconductors and advances predictive \emph{ab initio} computational approaches. We discuss possible extensions of our work. 
\end{abstract}

\maketitle

\section{Introduction}
First principles approaches based on density functional theory (DFT) and solution of the Boltzmann transport equation (BTE) have been widely used for many years now to give parameters-free determination of phonon and electron transport properties. In the typical treatment, phonons (electrons) are taken to be in equilibrium when solving the electron (phonon) BTE. This is a generalization of what is known as Blochsche Annahme (Bloch's Assumption) \cite{bloch1930}. This assumption was called into question by Peierls in 1930 \cite{peierls1930theorie}, who noted that the interactions between electrons and phonons causes each species to drag the other out of equilibrium. The theory of mutual drag of electrons and phonons was first formulated by L. E. Gurevich in 1946. The references to the original Russian articles are given in a review of the subject in Ref. \cite{gurevich1989electron}. In 1953, Frederikse found experimental evidence of the effect of phonon drag on the Seebeck coefficient of germanium \cite{fred1953}. This was followed by Seebeck coefficient measurements by Geballe and Hull in 1954 \cite{geballe1954seebeck} on germanium and in 1955 \cite{geballe1955seebeck} on silicon. In these measurements, a low temperature minimum of the absolute value of the Seebeck coefficient was observed. Conceptually, it was understood that phonons driven out of equilibrium by the applied temperature gradient induced an added electron current across the sample through electron-phonon coupling which in turn established a larger Seebeck voltage. Low temperatures were required to enter a regime where the scattering rates of the anharmonic phonon-phonon processes, which drive the phonon system toward equilibrium, became smaller than the phonon-electron scattering rates, thus helping set up a momentum flow between the electron and the phonon systems. Simple theories to explain the observed behavior were proposed by Frederikse in 1953 \cite{fred1953} and Conyers Herring in 1954 \cite{herring1954theory}. However, their theories explicitly violate the Kelvin-Onsager relation \cite{sond1956}. Since then, the drag effect has been observed in ZnO and CdS \cite{hutson1961piezoelectric}, Mg$_{2}$Si crystals \cite{heller1962seebeck}, and gold and platinum \cite{huebener1966effect}, among others. In recent years, computational approaches have been employed to study the drag phenomenon. In Ref. \cite{mahan2014seebeck}, Mahan, Lindsay, and Broido combined analytic models for the electron-phonon interaction with Rode's iterative BTE scheme \cite{rode1970electron} to a partially decoupled electron BTE. The phonon drag was captured by first principles calculations of the phonon-phonon interaction within the relaxation time approximation (RTA), and it was assumed that the non-equilibirum distribution of the electronic system does not affect the phonon system. This method captured the observed temperature dependence of the Seebeck coefficient in silicon \cite{geballe1955seebeck}. Zhou et. al. also studied the phonon drag effect in silicon by partially decoupling the electron and phonon BTEs \cite{zhou2015ab}. In their work the electron-phonon and phonon-phonon matrix elements were calculated from first principles. A first principles approach was also used by Fiorentini and Bonini in Ref. \cite{fiorentini2016thermo} to study silicon, and by Macheda and Bonini in Ref. \cite{macheda2018mag} to study p-doped diamond. In both these cases, a partial decoupling was implemented. To our knowlege, a full solution of the coupled electron and phonon BTEs has not been accomplished. In this work, we have fully solved the coupled electron-phonon BTEs, which enables us to study the mutual drag of electrons and phonons. Furthermore, it allows us to test how well the Kelvin-Onsager relations are satisfied in the meaningful low temperature regime where both electrons and phonons are substantially out of equilibrium. To test our approach, the coupled BTEs are solved using our in-house code for n-doped gallium arsenide (GaAs), and we show how this full solution accurately predicts the temperature dependence of the GaAs Seebeck coefficient, while the approach where phonons are forced to remain in equilibrium fails spectacularly to capture the observed behavior. We also show that phonon drag has a large effect on the mobility, but electron drag has only a small effect on the lattice thermal conductivity.

\section{Coupled transport of electron and phonons}
In this section we will formulate the coupled electron-phonon (e-ph) Boltzmann transport equations (BTE). Wherever applicable we will use the following mode denotations: $\lambda \equiv (s,\qq)$, where $s$ is the phonon polarization, and $\qq$ is the phonon wave vector; and $\nu \equiv (m,\kk)$, where $m$ is the electron electron band, and $\kk$ is the wave vector. We assume that electrons and phonons are well-defined quasiparticles and that a semiclassical description of their transport is valid. There are two coupled BTEs, one for the phonons and the other for the electrons:
\begin{align}\label{eq:masterphbte}
\bracom{\ddt n_{\lambda}}{field} &= -\bracom{\ddt n_{\lambda}}{coll},
\end{align}
\begin{align}\label{eq:masterebte}
\bracom{\ddt f_{\nu}}{field} &= -\bracom{\ddt f_{\nu}}{coll}, 
\end{align}
where $n_{\lambda}$ and $f_{\nu}$ are the non-equilibrium distribution functions of the phonon and the electron systems, respectively. These equations describe a steady state situation where the diffusion of particles due to an external field is balanced by the collisions the particles experience. Since the e-ph interaction drives both systems out of equilibrium, their transport is coupled. This implies that even if an external field does not directly cause drift of a species, there will still be a response from that species via its interaction with another species that couples directly to the field. The phenomenon of e-ph coupling induced drift is known as the drag effect or Gurevich effect after L. E. Gurevich who studied it extensively \cite{gurevich1989electron}.

We assume that the external fields (electric field, $\mb{E}$ and temperature gradient, $\nabla T$) are weak and, as such, the system of particles will repond to it by going out of equilibrium by an amount that is proportional to the relevant field. For the electron system the non-equilibrium distribution function is given by
\begin{align}
f_{\nu} &\approx f^{0}_{\nu} - \dfrac{1}{\beta}\partial_{\epsilon_{\nu}}f^{0}_{\nu}\Psi_{\nu} \nonumber \\
&= f^{0}_{\nu}\left[1 + (1-f^{0}_{\nu})\Psi_{\nu}\right],
\end{align}
where $f^{0}_{\nu}$ is the equilibrium (Fermi-Dirac) electron distribution; $\epsilon_{\nu}$ is electron energy; $\Psi_{\nu}$ is the deviation function; and $\beta \equiv (k_{B}T)^{-1}$, where $k_{B}$ is the Boltzmann constant and $T$ is the temperature. 

The deviation function is a linear response to the external fields and is given by
\begin{equation}\label{eq:eldev}
\Psi_{\nu} = -\beta\nabla T\cdot \mb{I}_{\nu} - \beta\mb{E}\cdot \mb{J}_{\nu},
\end{equation}
where $\nabla T$ and $\mb{E}$ are the temperature gradient and electric field, and $\mb{I}_{\nu}$ and $\mb{J}_{\nu}$ are the corresponding response functions.

Similarly, the phonon non-equilibrium distribution is given by
\begin{align}
n_{\lambda} &\approx n^{0}_{\lambda} - \dfrac{1}{\beta}\partial_{\hbar\omega_{\lambda}}n^{0}_{\lambda}\Phi_{\lambda} \nonumber \\
&= n^{0}_{\lambda}\left[1 + (1+n^{0}_{\lambda})\Phi_{\lambda}\right],
\end{align}
where $n^{0}_{\lambda}$ is the equilibrium (Bose-Einstein) phonon distribution, $\omega_{\lambda}$ is the angular frequency, and $\Phi_{\lambda}$ is the deviation function. In this case the deviation function is given by
\begin{equation}\label{eq:phdev}
\Phi_{\lambda} = -\beta\nabla T\cdot \mb{F}_{\lambda} - \beta\mb{E}\cdot \mb{G}_{\lambda},
\end{equation}
where $\mb{F}_{\lambda}$ and $\mb{G}_{\lambda}$ are the responses to the temperature gradient and electric field, respectively. Note here that phonons do not directly couple to the electric field. However, because of the e-ph interaction, there is an indirect response of the phonons to the electric field. Thus $\mb{G}_{\lambda}$ is entirely capturing the drag effect, whereas $\mb{I}_{\nu}$, $\mb{J}_{\nu}$ and $\mb{F}_{\lambda}$ capture both the direct field effect and the drag effect.
\subsection{Field and collision terms}
The electron drift term is given by
\begin{align}
\bracom{\ddt f_{\nu}}{field} &= \dfrac{e}{\hbar}\mb{E}\cdot\nabla_{\kk}\fdsub{\nu} - \velsub{\nu}\cdot\nabla T\partial_{T}\fdsub{\nu} \nonumber \\
&= \beta\fdsub{\nu}(1-\fdsub{\nu})\left[ -e\mb{E} - \dfrac{\ensub{\nu}-\mu}{T}\nabla T \right]\cdot\velsub{\nu},
\end{align}
where $e$ is the magnitude of the bare electron charge, $\velsub{\nu}$ is the electron velocity, and $\mu$ is the chemical potential corresponding to a given electron concentration.

The drift term for the phonon system is
\begin{align}
\bracom{\ddt n_{\lambda}}{field} &= -\velsub{\lambda}\cdot\nabla T\partial_{T}\besub{\lambda} \nonumber \\
&= -\beta\besub{\lambda}(1+\besub{\lambda})\dfrac{\hbar\omega_{\lambda}}{T}\nabla T\cdot\velsub{\lambda},
\end{align}
where $\velsub{\lambda}$ is the phonon group velocity. As mentioned before, phonons do not directly respond to the electric field, which is why in the expression above there is no $\mb{E}$ dependent term.

Now, we discuss the collision terms. We relegate the consideration of electron-electron collisions for some later time and assume that the intrinsic electrical resistance arises from the lowest order e-ph interaction.

	

We first calculate the scattering probabilities corresponding to the phonon absorption and emission phonon processes: 
\begin{widetext}
\begin{align}\label{eq:ex}
\stirling{X^{+}_{\nu\nu'\lambda}}{X^{-}_{\nu\nu'\lambda}} = \dfrac{2\pi}{\hbar}\gsq \stirling{ \fdsub{m\kk}(1-\fdsub{n[\kk+\qq]})\besub{s\qq}\delta(\elen{n[\kk+\qq]}-\elen{m\kk}-\phen{s\qq}) } { \fdsub{m\kk}(1-\fdsub{n[\kk+\qq]})(1+\besub{s-\qq})\delta(\elen{n[\kk+\qq]}-\elen{m\kk}+\phen{s-\qq}) }
\end{align}
\end{widetext}
where we have introduced the notation $[\kk+\qq] \equiv (\kk+\qq)\text{ modulo }\GG$, where $\GG$ is the reciprocal lattice vector. From now on we will use the fact that $\besub{-\qq} = \besub{\qq}$ since $\omega_{-\qq} = \omega_{\qq}$, from time reversal symmetry. The electron-phonon interaction matrix element is given by \cite{giustino2007electron}
\labeq{
	g^{smn}_{\mb{k}\qq} = \sqrt{\dfrac{\hbar}{2M\omega_{s\qq}}}\bra{\zeta_{n[\mb{k}+\qq]}}\nabla_{s\qq}V_{\text{SCF}}\ket{\zeta_{m\mb{k}}},
}
where $\zeta_{m\mb{k}}$ is the electronic (Kohn-Sham) state, $M$ is a convenient reference mass and $V_{\text{SCF}}$ is the self-consistent potential.

Then, using Fermi's golden rule followed by linearization in the response functions, we find for the electronic case:
\begin{widetext}
\begin{align}
\bracom{\ddt f_{m\kk}}{collision}^{\text{e-ph}} = \beta \sum_{ns\qq}&\Bigg[ X^{+}_{\nu\nu'\lambda}\stirling{\Isub{m\kk}-\Isub{n[\kk+\qq]}+\Fsub{s\qq}}{\Jsub{m\kk}-\Jsub{n[\kk+\qq]}+\Gsub{s\qq}}                                                                                     + X^{-}_{\nu\nu'\lambda}\stirling{\Isub{m\kk}-\Isub{n[\kk+\qq]}-\Fsub{s-\qq}}{\Jsub{m\kk}-\Jsub{n[\kk+\qq]}-\Gsub{s-\qq}} \Bigg]\cdot\stirling{\nabla T}{\mb{E}},
\end{align} 
\end{widetext}
where the top (bottom) line inside the braces corresponds to the temperature gradient (electric) field.

For the phonon system there are two intrinsic scattering sources -- anharmonic phonon scattering, and phonon scattering from electrons. The three-phonon scattering probabilities are \cite{ziman1960electrons}
\begin{align}\label{eq:phprocs}
W^{\pm}_{\lam\lam'\lam''} =&\dfrac{\pi\hbar}{4}\dfrac{|V^{\pm}_{\lam\lam'\lam''}|^{2}}{\omega_{\lam}\omega_{\lam'}\omega_{\lam''}}(\besub{\lam}+1)\left(\besub{\lam'}+\dfrac{1}{2} \pm \dfrac{1}{2}\right)\besub{\lam''}\nonumber\\
&\times\delta(\omega_{\lam}\pm\omega_{\lam'}-\omega_{\lam''}),
\end{align}
where $V^{\pm}_{\lam\lam'\lam''}$ is the three-phonon scattering matrix element, and $\qq'' = (\qq \pm \qq')\text{ modulo } \GG$ for the plus and minus processes, respectively.

For the $\pm$ processes defined in Eq. \ref{eq:phprocs} the three-phonon scattering matrix elements are given by \cite{wu2014}
\begin{equation}
V^{\pm}_{\lambda\lambda'\lambda''} = \sum_{<i>jk}\sum_{\alpha\beta\gamma}\Psi^{\alpha\beta\gamma}_{ijk}\dfrac{e^{i\alpha}_{s,\mathbf{q}}e^{j\beta}_{s',\pm\mathbf{q}'}e^{k\gamma}_{s'',-\mathbf{q}''}}{\sqrt{m_{i}m_{j}m_{k}}},
\end{equation}
where $i,j,k$ label atoms in the supercell with $<>$ symbolizing restricted sum over the primitive cell, $m_{i}$ is the mass of atom $i$, $e^{i\alpha}_{s,\mathbf{q}}$ is the $\alpha$ Cartesian component of the phonon eigenvector associated with polarization $s$ and wavevector $\mathbf{q}$, and $\Psi^{\alpha\beta\gamma}_{ijk} = \dfrac{\partial^{3}U}{\partial r^{\alpha}_{i}\partial r^{\beta}_{j}\partial r^{\gamma}_{k}}$ is the third-order anharmonic force constant where $r^{\alpha}_{i}$ is the displacement of atom $i$ in the Cartesian direction $\alpha$ calculated from the crystal potential energy $U$.\\
\\
Similarly, we calculate the ph-e scattering probability as
\begin{align}\label{eq:y}
Y_{\lam mn\kk} = &\dfrac{2\pi}{\hbar}\gsq\fdsub{m\kk}(1-\fdsub{n[\kk+\qq]})\besub{\lam}\nonumber\\
&\times\delta(\elen{n[\kk+\qq]}-\elen{m\kk}-\phen{\lam}).
\end{align}
We can now write down the two intrinsic phonon collision terms as follows:
\begin{widetext}
\begin{align}
\bracom{\partial_{t}n_{\lam}}{collision}^{\text{3ph}} = \beta\sum_{\lam'\lam''}&\Bigg[W^{+}_{\lam\lam'\lam''}\stirling{\Fsub{\lam}+\Fsub{\lam'}-\Fsub{\lam''}}{\Gsub{\lam}+\Gsub{\lam'}-\Gsub{\lam''}}+ \dfrac{1}{2}W^{-}_{\lam\lam'\lam''}\stirling{\Fsub{\lam}-\Fsub{\lam'}-\Fsub{\lam''}}{\Gsub{\lam}-\Gsub{\lam'}-\Gsub{\lam''}}\Bigg]\cdot\stirling{\nabla T}{\mb{E}}
\end{align}

\begin{align}
\bracom{\partial_{t}n_{\lam}}{collision}^{\text{ph-e}} = 2\beta\sum_{mnk}Y_{\lam mn\kk}\stirling{\Isub{m\kk}-\Isub{n[\kk+\qq]}+\Fsub{\lam}}{\Jsub{m\kk}-\Jsub{n[\kk+\qq]}+\Gsub{\lam}}\cdot\stirling{\nabla T}{\mb{E}},
\end{align}
\end{widetext}
where the leading $2$ in the last equation is due to the spin degree of freedom of the electron in band $m$ and wavevector $\kk$.

\subsection{Coupled BTEs}
Two coupled pairs of BTEs corresponding to the two driving fields are obtained by plugging in the relevant collision and field terms into Eqs. \eqref{eq:masterphbte}, \eqref{eq:masterebte}.
\subsubsection{Phonon response to temperature gradient}
We first define the following objects:
\labalign{
	Q_{\lam} = &\sum_{\lam'\lam''}\left(W^{+}_{\lam\lam'\lam''}+\dfrac{1}{2}W^{-}_{\lam\lam'\lam''}\right) \nonumber \\ 
	&+ 2\sum_{mn\kk}Y_{\lam mn\kk}  + \sum_{j\in\text{channels}}Q^{j}_{\lam},
}
\labalign{
	\Delta\Fsub{\text{S},\lam} = \dfrac{1}{Q_{\lam}}\sum_{\lam'\lam''}&\big[ W^{+}_{\lam\lam'\lam''}\left( \Fsub{\lam''}-\Fsub{\lam'} \right) \nonumber\\
	&+ \dfrac{1}{2}W^{-}_{\lam\lam'\lam''}\left( \Fsub{\lam''}+\Fsub{\lam'} \right) \big],
}
\labeq{
	\Delta\Fsub{\text{D},\lam} = \dfrac{2}{Q_{\lam}}\sum_{mn\kk}Y_{\lam mn\kk}\left(\Isub{n[\kk+\qq]}-\Isub{m\kk}\right),
}
\labeq{\label{eq:Tpbte0}
	\Fsub{\lam}^{0} = \dfrac{\phen{\lam}\velsub{\lam}\besub{\lam}(\besub{\lam}+1)}{Q_{\lam}T},
}
where subscripts S and D stand for ``self'' (functional of own deviation function) and ``drag'' (functional of the other species' deviation function). In the first equation provision has been made to add additional scattering channels like defect scattering. The additional single phonon scattering in channel $j$ is denoted $Q^{j}_{\lambda}$.

In terms of these, the phonon BTE due to the $\nabla T$ field becomes
\labeq{\label{eq:Tphbte}
	\Fsub{\lam} = \Fsub{\lam}^{0} + \Delta\Fsub{\text{S},\lam} + \Delta\Fsub{\text{D},\lam}
}
where $i$ in brackets denotes the iteration number.

We identify the phonon relaxation rate from the function $Q_{\lam}$ in the following way
\labeq{\label{eq:phrta}
	W^{\text{ph,RTA}}_{\lam} = \dfrac{Q_{\lam}}{\besub{\lam}(\besub{\lam}+1)}.
}
%
\\
Note here that, if one assumes that during the e-ph scattering the electrons remain in equilibrium ($I_{\nu} = 0$), then $\Delta \Fsub{\text{D},\lam} = 0$. This is the typical approximation used in treating the effect of ph-e scattering on the thermal conductivity \cite{liao2015,protik2017,wangsic,li2018fermi}. In those works, ph-e scattering was also treated in the RTA, while ph-ph scattering was treated in full.

\subsubsection{Phonon response to electric field}
As mentioned before, the electric field does not directly cause a diffusion of phonons. Instead, the effect of the electric field is felt by the phonon system due to the e-ph interaction, i.e., via the drag effect. We first define the following self and drag functions:
\labalign{
	\Delta\Gsub{\text{S},\lam} = \dfrac{1}{Q_{\lam}}\sum_{\lam'\lam''}\big[ &W^{+}_{\lam\lam'\lam''}\left( \Gsub{\lam''}-\Gsub{\lam'} \right)\nonumber\\
	& + \dfrac{1}{2}W^{-}_{\lam\lam'\lam''}\left( \Gsub{\lam''}+\Gsub{\lam'} \right) \big],
}
\labeq{
	\Delta\Gsub{\text{D},\lam} = \dfrac{2}{Q_{\lam}}\sum_{mn\kk}Y_{\lam mn\kk}\left(\Jsub{n[\kk+\qq]}-\Jsub{m\kk}\right).
}

In terms of these, the phonon BTE due to the electric field becomes
\labeq{\label{eq:Ephbte}
	\Gsub{\lam} = \Delta\Gsub{\text{S},\lam} + \Delta\Gsub{\text{D},\lam}.
}

\subsubsection{Electron response to temperature gradient}
We start by defining the total scattering probability due to the e-ph interaction as follows:
\labeq{\label{eq:er}
	R_{\nu} = \sum_{n\lam} \left(X^{+}_{\nu n[\kk+\qq]\lam} + X^{-}_{\nu n[\kk+\qq]\lam}\right) + \sum_{j\in\text{channels}}R^{j}_{\nu},
}
where $R^{j}_{\nu}$ denotes some additional, single electron scattering term in channel $j$.

Next, we define the self and drag functions:
\labeq{
	\Delta \Isub{\text{S},\nu} = \dfrac{1}{R_{\nu}}\sum_{ns\qq} \Isub{n[\kk+\qq]}\left(X^{+}_{\nu n[\kk+\qq]\lam} + X^{-}_{\nu n[\kk+\qq]\lam}\right),
}
\labeq{\label{eq:TebteD}
	\Delta \Isub{\text{D},\nu} = \dfrac{1}{R_{\nu}}\sum_{ns\qq} \left(X^{-}_{\nu n[\kk+\qq]\lam}\Fsub{s-\qq} - X^{+}_{\nu n[\kk+\qq]\lam}\Fsub{s\qq}\right).
}

Lastly, we define
\labeq{
	\Isub{\nu}^{0} = \dfrac{\elen{\nu}-\mu}{R_{\nu}T}\fdsub{\nu}\left(1-\fdsub{\nu}\right)\velsub{\nu}.
}

In terms of these functions, the $\nabla T$ response electron BTE can be written as
\labeq{\label{eq:Tebte}
	\Isub{\nu} = \Isub{\nu}^{0} + \Delta\Isub{\text{S},\nu} + \Delta\Isub{\text{D},\nu}.
}

We can identify the relaxation rate as
\labeq{\label{eq:erta}
	W^{\text{e,RTA}}_{\nu} = \dfrac{R_{\nu}}{\fdsub{\nu}\left(1-\fdsub{\nu}\right)}.
}
\\
Note that Eqs. \eqref{eq:Tphbte} and \eqref{eq:Tebte} are coupled because of the presence of the drag terms. In general, these two equations must be solved together to self-consistency in order to capture the effect of drag on the transport coefficients of the two systems.

\subsubsection{Electron response to electric field}
We proceed as before and define the self and drag functions:
\labeq{
	\Delta \Jsub{\text{S},\nu} = \dfrac{1}{R_{\nu}}\sum_{ns\qq} \Jsub{n[\kk+\qq]} \left(X^{+}_{\nu n[\kk+\qq]\lam} + X^{-}_{\nu n[\kk+\qq]\lam}\right),
}
\labeq{
	\Delta \Jsub{\text{D},\nu} = \dfrac{1}{R_{\nu}}\sum_{ns\qq} \left(X^{-}_{\nu n[\kk+\qq]\lam}\Gsub{s-\qq} - X^{+}_{\nu n[\kk+\qq]\lam}\Gsub{s\qq}\right).
}

Then we define
\labeq{
	\Jsub{\nu}^{0} = \dfrac{e}{R_{\nu}}\fdsub{\nu}\left(1-\fdsub{\nu}\right)\velsub{\nu}.
}

And in terms of these we find the $\mb{E}$ response electron BTE:
\labeq{\label{eq:Eebte}
	\Jsub{\nu} = \Jsub{\nu}^{0} + \Delta\Jsub{\text{S},\nu} + \Delta\Jsub{\text{D},\nu}.
}
\\
Having derived the four coupled BTEs, we now set up an iterative scheme to solve them.
\subsubsection{Iteration scheme}
\begin{figure}
	\centering
	\includegraphics[width=1.0\linewidth]{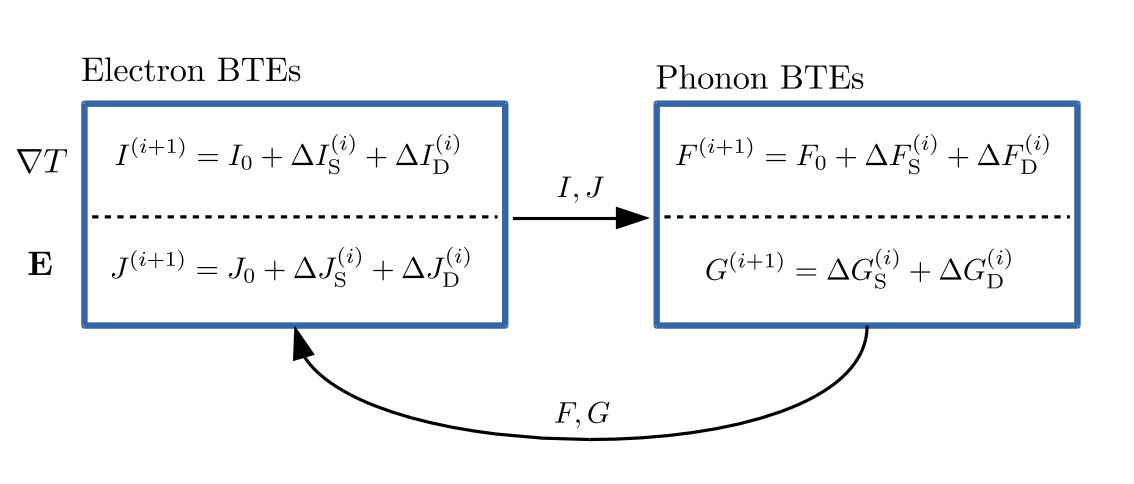}
	\caption{An iterative scheme for solving the coupled electron-phonon BTEs.}
	\label{fig:flow}
\end{figure}

In Fig \ref{fig:flow} we outline the iterative scheme that we have developed to solve Eqs. \ref{eq:Tphbte} and \ref{eq:Tebte} for an applied temperature gradient, and Eqs. \ref{eq:Ephbte} and \ref{eq:Eebte} for an applied electric field. For a given field, we start with the simultaneous evaluation of the zeroth order response functions, i.e. setting all the self and drag terms to be zero. This is nothing but the RTA. This approximation feeds into the calculation of the self and the drag functions of the next iteration to give the first order response functions. This process is continued until the response functions of both the electronic and the phonon systems converge.

Let us now take note of some properties of this formulation. First, the $\nabla T$ and $\mb{E}$ equations completely decouple. Second, the formulation is consistent with what is expected if the e-ph interaction is switched off: For the electronic system we find that $\mb{I}$ and $\mb{J}$ blow up if no other electron scattering mechanism is present. For the phonon system we find that $\mb{G} = 0$ at all times owing to the lack of the drag effect. Also, as mentioned before, the $\mb{F}$ equation reduces to the well-known phonon BTE \cite{wu2014,peierls1929kinetischen} involving only ph-ph interactions. 

\subsubsection{Transport coefficients}
In this section we derive the complete set of transport coefficients that we can calculate from the solutions of the coupled BTEs described above. We begin with the carrier transport. In the linear response regime, the charge current density is
\labeq{
	\mb{\mathcal{J}}_\text{c} = L_{11}\mb{E} + L_{12}(-\mb{\nabla} T),
}
where, $L_{ij}$ is a rank-$2$ electronic transport tensor.
\\
This current density is given by
\labalign{
	\mb{\mathcal{J}}_\text{c} &= -\dfrac{2e}{V}\sum_{\nu}\mb{v}_{\nu}f_{\nu} \nonumber\\
		&= -\dfrac{2e}{V}\sum_{\nu}\mb{v}_{\nu}f^{0}_{\nu} -\dfrac{2e}{V}\sum_{\nu}\mb{v}_{\nu}f^{0}_{\nu}(1-f^{0}_{\nu})\Psi_{\nu}.
}
\\
The first term of the second line vanishes due to time-reversal symmetry. Using Eq. \ref{eq:eldev} and comparing the two expressions above, we obtain the transport tensors
\labeq{
	\stirling{L_{11}}{L_{12}} = \dfrac{2e}{Vk_{B}T}\sum_{\nu}f^{0}_{\nu}(1-f^{0}_{\nu})\mb{v}_{\nu}\otimes\stirling{\mb{J}_{\nu}}{-\mb{I}_{\nu}}.
}
\\
We can identify these with the familiar transport coefficients: the carrier conductivity tensor $\sigma^{\alpha\beta} = L_{11}^{\alpha\beta}$ carrying the units $\Omega^{-1}$m$^{-1}$, and the conductivity times Seebeck coeffiecient tensor $[\sigma S]^{\alpha\beta} = L_{12}^{\alpha\beta}$ in units of Am$^{-1}$K$^{-1}$.\\

Next, we look at the electronic heat current density
\labalign{
	\mb{\mathcal{J}}_\text{el,h} &= L_{21}\mb{E} + L_{22}(-\mb{\nabla} T)\nonumber\\
		&= \dfrac{2}{V}\sum_{\nu}(\epsilon_{\nu}-\mu)\mb{v}_{\nu}f^{0}_{\nu}(1-f^{0}_{\nu})\Psi_{\nu}.
}
\\
From the above we get
\labeq{
	\stirling{L_{21}}{L_{22}} = -\dfrac{2}{Vk_{B}T}\sum_{\nu}(\epsilon_{\nu}-\mu)f^{0}_{\nu}(1-f^{0}_{\nu})\mb{v}_{\nu}\otimes\stirling{\mb{J}_{\nu}}{-\mb{I}_{\nu}}.
}
\\
The electronic thermal response to an electric field is usually denoted $\alpha_{\text{el}}^{\alpha\beta} = L^{\alpha\beta}_{21}$ in units of Am$^{-1}$. And the thermal response to the temperature gradient is the $\mb{E}=0$ thermal conductivity $\kappa_{0,\text{el}}^{\alpha\beta} = L^{\alpha\beta}_{22}$ in units of Wm$^{-1}$K$^{-1}$. The electronic thermal conductivity at zero electric current is given by
\labeq{
	\kappa_{\text{el}}^{\alpha\beta} = \kappa_{\text{0,el}}^{\alpha\beta} - \alpha_{\text{el}}^{\alpha\beta}S^{\alpha\beta}.
}

Now, we consider the phonon thermal current density
\labalign{
	\mb{\mathcal{J}}_\text{ph,h} &= K_{21}\mb{E} + K_{22}(-\mb{\nabla} T)\nonumber\\
	&= \dfrac{1}{V}\sum_{\lam}\hbar\omega_{\lam}\mb{v}_{\lam}n^{0}_{\lam}(1+n^{0}_{\lam})\Phi_{\lam}.
}
where, $K_{ij}$ is a rank-$2$ phonon transport tensor. Note that unlike the electronic case, $K_{11}$ and $K_{12}$ tensors are zero as phonons do not carry a charge.\\
\\
Using Eq. \ref{eq:phdev} we obtain from the above
\labeq{\label{eq:phtranscoeffs}
	\stirling{K_{21}}{K_{22}} = \dfrac{1}{Vk_{B}T}\sum_{\lam}\hbar\omega_{\lam}n^{0}_{\lam}(1+n^{0}_{\lam})\mb{v}_{\lam}\otimes\stirling{-\mb{G}_{\lam}}{\mb{F}_{\lam}}.
}
\\
We identify $\alpha_{\text{ph}}^{\alpha\beta} = K^{\alpha\beta}_{21}$, which is to be added to its electronic counterpart to obtain
\labeq{
	\alpha_{\text{total}}^{\alpha\beta} = \alpha_{\text{el}}^{\alpha\beta} + \alpha_{\text{ph}}^{\alpha\beta}.
}
\\
Also, we identify the phonon thermal conductivity as $\kappa_{\text{ph}}^{\alpha\beta} = K_{22}^{\alpha\beta}$, which again is to be added to the total electronic contribution.\\
\\
In total there are three independent transport coefficients that we can compute, as two of the coefficients are related by a Kelvin-Onsager relation \cite{sond1956}
\labeq{\label{eq:kelvin}
	\dfrac{\alpha_{\text{total}}}{T} = S\sigma.
}
\\
This theorem holds for time-reversal invariant systems and when the linear response regime is valid. As such, it provides a strong check for the self-consistency of our formulation.

\section{Model and computational details}
We apply the above calculational scheme to the common semiconductor, GaAs. GaAs has a measured direct band gap of around $1.4$ eV at room temperature \cite{gaascollection}. Using the LDA exchange-correlation functional, however, we get around $0.3$ eV. This leads to a significant increase in the conduction band mass \cite{sofo1994optimum}. While it is possible \cite{liu2017first} to reproduce the actual band gap and band mass for this material by including many-body corrections, such calculations are computationally expensive. For this work, we simply assume a single parabolic, isotropic conduction band of mass $0.067m_{e}$ \cite{gaascollection}. This approximation should work well for GaAs for moderate electron densities ($<$ $10^{19}$cm$^{-3}$) where only the bottom of the conduction band is relevant for transport. Doping is simulated by shifting the chemical potential for the system. The phonon dispersions are calculated \emph{ab initio}.

Since the electronic energy scale is of the order of eV while the largest phonon energy scale is usually tens of meV, a very dense electronic wavevector mesh is required to calculate the electron-phonon phase space for scattering, the matrix elements, and to do the Brillouin zone sums to obtain electronic transport coefficients. However, calculation of electron-phonon matrix elements on such a dense mesh is computationally expensive. The matrix elements also have to be saved for reuse during the iterative BTE solution. To keep the computation tractable, we take two strategies: 1) We use simple, physical models for the electron-phonon interaction. This way we save a massive amount of computational time and storage that would otherwise go toward the calculation of the matrix elements. 2) We use a Fermi window of $0.5$ eV from the bottom of the conduction band beyond which we do not compute any electronic quantities. Since the contributions to the electronic transport coefficients come mostly from around the chemical potential and since our chemical potential will remain close to the conduction band minimum for the carrier concentrations we will consider in this work, choosing such a Fermi window cutoff is justified. The simple model electron-phonon interaction channels are the acoustic deformation potential (ADP), piezoelectric potential (PZ) for LA phonons, and polar optic phonon (POP) scattering. Details about the origins of these models can be found in Ref. \cite{mahan2011}. Briefly, long wavelength acoustic phonons produce lattice dilation and contraction. This local change in the lattice constant causes a time periodic shift in the electronic bands. The proportionality constant of the shift in the band energy and the change in the lattice constant is called the ADP. A related phenomenon is the PZ scattering, which happens in polar materials without a center of inversion. In this case, acoustic vibrational modes create a long range electric field that scatters electrons. The constant of proportionality of the local strain and the electric field is the PZ constant. Lastly, POP is an interaction prevalent in polar materials. Long wavelength LO phonon vibrational modes create oscillating dipoles in every unit cell, which leads to a long range electric field that scatters electrons. The matrix element for this scattering can be computed by considering the lowest order Feynmann diagrams for both the screened Coulomb and phonon mediated electron-electron interactions. This is also known as the Fr\"{o}hlich model \cite{mahan2011}. The matrix elements for these processes are given by \cite{nag}
\labalign{
	g^{\text{ADP}}_{\lam} = \left(\dfrac{\hbar}{2V\rho\omega_{\lam}}\right)^{1/2}Aq, 
}
\labalign{
	g^{\text{PZ}}_{\lam} = \left(\dfrac{ \hbar e^{2} e_{\text{PZ}}^{2} }{2V\rho c_{\text{PZ}}\varepsilon_{0}^{2}\kappa_{\infty}^{2}q}\right)^{1/2} \left( \dfrac{q^{2}}{q^{2} + q_{\text{TF}}^{2}} \right), \text{and}
}
\labalign{
	g^{\text{POP}}_{\lam} = \left( \dfrac{e^{2}\hbar\omega_{\lam}}{2\varepsilon_{0}V} \right)^{1/2}\dfrac{1}{q}\left(\dfrac{1}{\kappa_{\infty}} - \dfrac{1}{\kappa_{0}}\right)^{1/2}\left( \dfrac{q^{2}}{q^{2} + q_{\text{TF}}^{2}} \right),
}
where $\rho$ is the density, $V$ is the crystal volume, $\varepsilon_{0}$ is the permittivity of free space, $\kappa_{0} = 13.1$ is the static dielectric constant, and $\kappa_{\infty} = 11.1$ is the high frequency dielectric constant. Here we have taken the PZ and POP interactions to be screened within a Thomas-Fermi model, and $q_{\text{TF}}$ is the static Thomas-Fermi screening wavevector. We use the high-frequency dielectric constant in the screening wavevector calculation. The ADP parameter is $A = 7$ eV \cite{rode1970electron}. The PZ contribution from all acoustic phonons is approximated as merged into a single acoustic branch, and from Ref. \cite{rode1970electron} the parameters are the averaged acoustic speed $c_{\text{PZ}} = 4030$ ms$^{-1}$ and the piezoelectric constant $e_{\text{PZ}} = 0.16$ Cm$^{-2}$.

Electron-electron interactions are ignored in this work and we provide two justifications why. First, the transport active electrons reside near the conduction band minimum at the $\Gamma$-point. As such, electron-electron Umklapp processes, which must involve at least one large magnitude wavevector, are completely negligible. Also, first principles calculations showed that electron-electron self-energies are smaller than electron-phonon self-energies \cite{liu2017first}. However, the electron-electron interaction could provide dynamical screening to other types of interactions in the system, which brings us to the second point: The inclusion of electron-electron interactions typically involves infinite order diagrammatics known as the random phase approximation (RPA) \cite{mahanbook}, which gives rise to additional bosonic excitations of the electron density. These are known as plasmons. The RPA dielectric function is both momentum and energy dependent, which adds considerable complexity over the static Thomas-Fermi screening model we currently use. Furthermore, the plasmons interact with the LO phonons, leading to the formation of hybrid coupled modes \cite{hauberfahy}. Also, the RPA gives rise to a region in the energy-momentum phase space known as an electron-hole pair excitation continuum (PEC). Inside this continuum, plasmons are strongly damped via their interaction with short-lived, bosonic electron-hole pairs, and as such do not maintain a sharp energy dispersion relation \cite{hauberfahy}. As the BTE is valid only for sharp modes the applicability of our current approach breaks down. In passing we comment on the various approaches attempted to take electron-electron interactions into account. In Ref. \cite{sanborn1995electron}, Sanborn used the plasmon-pole approximation which captures the spectral weight of the non-sharp electron-hole pair excitations into an effective plasmon mode going through the PEC. Following this the BTE was applied. In Ref. \cite{caruso2016theory}, Caruso and Giustino calculated the self-energy relaxation rate from first principles for plasmon-electron scattering. In that work, the relaxation time approximation was used, so the full solution of the BTE was not attempted. In Ref. \cite{hauberfahy}, Hauber and Fahy applied the BTE to both the sharp plasmons and the non-sharp pair excitations. In their work a coupled system of BTEs -- one for the LO phonon-plasmon coupled modes, the other for the electronic system -- is solved using simple analytical models. We implemented their method and found that the questionable use of the BTE on low energy, non-sharp, pair excitations leads to a large mobility gain in GaAs and GaN. In Ref. \cite{wu2019}, Wu et. al. used much of the methodology developed by Hauber and Fahy, but did not allow the bosonic coupled modes to go out of equilibrium, while retaining the use of the RPA. We do not dwell on this issue further, but remark that a computationally feasible, rigorous treatment of coupled electron-phonon transport with dynamical screening effects remains an open problem worthy of deeper investigation.

Starting with the opensource, uncoupled phonon BTE solver, \texttt{ShengBTE} \cite{wu2014} codebase, we built our coupled e-ph BTE solver - \texttt{elphBolt} (\underline{el}ectron-\underline{ph}onon \underline{Bol}tzmann \underline{t}ransport). Our code is capable of using DFT based electronic band structure and phonon dispersion. Also, e-ph matrix elements calculated from first principles methods can be read in for a completely parameters-free calculation. We relegate a fully \emph{ab initio} calculation to the future. In this work we made simplifying assumptions regarding the e-ph matrix elements. Here we employed a dual mesh approach, with an ultrafine electron wavevector ($\kk$) mesh and a relatively coarser phonon wavevector ($\qq$) mesh. When needed, phonon quantities from the coarse mesh are interpolated on to the fine mesh. There are three run levels of \texttt{elphBolt}. In run level 1, the irreducible coarse electronic wavevector mesh is generated. The user then calculates the electronic band on this mesh and provides it as an input in runlevel 2. In this second step, \texttt{elphBolt} performs a mesh refinement and applies a Fermi surface thickness. This generates an ultrafine mesh restricted by the specified Fermi surface thickness. The user then calculates the electronic bands on this refined mesh and provides it as an input for run level 3. In this last step, phonon dispersion and ph-ph matrix element calculations are performed. Also, the model e-ph and ph-e matrix elements are calculated. All matrix elements are saved for future use. Finally, the coupled BTEs are solved iteratively for the specified driving field. The computationally expensive parts of the code - phonons, phonon-phonon matrix elements and BTE iterations - are parallelized using MPI. Coupling between the BTEs can be switched off if the user wants to solve for transport without the drag effect. The phonon-phonon matrix elements are calculated \emph{ab initio}. The analytic tetrahedron method \cite{lambin1984computation} is used to evaluate the energy conserving delta functions appearing in all the matrix element calculations. We use the Quantum Espresso suite \cite{giannozzi2017advanced} for our DFT needs.

The bottleneck of the program is the full \emph{ab initio} calculation of the ph-ph matrix elements. On the $60\times60\times60$ $\qq$-mesh, this calculation requires nearly $3000$ cpu-hours. This produces $55$ GB of data that are saved on the disk for later read-in during the iteration process. For contrast, a $90\times90\times90$ $\qq$-mesh ph-ph matrix element calculations requires nearly $14000$ cpu-hours and $390$ GB disk space, rendering it a monumental computational task. In comparison, the e-ph matrix elements require much smaller time and memory owing to the usage of the effective Fermi surface thickness and analytic models. Once the matrix elements have been calculated, a solution of the coupled BTEs at a given carrier concentration takes nearly $3000$ cpu-hours to finish on the $60\times60\times60$ $\qq$-, $600\times600\times600$ $\kk$-mesh. This number is nearly $14000$ cpu-hours for the $90\times90\times90$ $\qq$-, $630\times630\times630$ $\kk$-mesh. For comparison, an uncoupled electron BTE (i.e. phonons taken to be in equilibrium) takes a few hundred cpu-hours on the $60\times60\times60$ $\qq$-, $600\times600\times600$ $\kk$-mesh.
\begin{table}
	\centering
	\begin{tabular}{c|c|c|c}
		mesh & $\kappa^{\text{el}}$+$\kappa^{\text{ph}}$ (W$^{-1}$m$^{-1}$K$^{-1}$) & $\sigma$ ($10^{6}$$\Omega^{-1}$m$^{-1}$) & $S$ ($\mu$V$^{-1}$K$^{-1}$)\\
		\hline
		$(30,300)$ & $10.11 + 1237$ & $11.67$ & $-160.0$\\
		$(45,450)$ & $9.93 + 1266$ & $16.60$ & $-123.7$\\
		$(60,600)$ & $9.94 + 1295$ & $20.81$ & $-141.9$\\
		$(90,630)$ & $10.20 + 1317$ & $25.18$ & $-134.0$\\
	\end{tabular}
	\caption{Transport coefficients calculated from the e-ph coupled BTEs solutions for both the temperature gradient and electric field for various mesh densities. The temperature is set at $50$ K, the lowest considered in this work and for which case the drag effect is the strongest, and the carrier concentration is set at a high $10^{18}$ cm$^{-3}$. The notation $(n,m)$ in first column denotes an $n^{3}$ $\qq$-mesh and an $m^{3}$ $\kk$-mesh. Data in the subsequent columns are respectively the electronic + lattice thermal conductivity, charge conductivity, and Seebeck coefficient.}
	\label{tab:conv}
\end{table}

The transport coefficients for various mesh densities are shown in Tab. \ref{tab:conv}. For the main calculations presented in this work we calculate the electronic quantities on a $600\times600\times600$ $\kk$-mesh, while the phonon quantities are calculated on a $60\times60\times60$ $\qq$-mesh. While this does not give tightly converged results for all the transport coefficients we look at low temperatures, given the large time and memory requirements of the matrix elements calculation and the coupled BTEs solutions, we find this to be best balance between economy and accuracy. Moreover, as we will show later, the current choice of mesh densities is sufficient to capture strong drag physics. 

\begin{table}[H]
	\centering
	\begin{tabular}{c|c|c|c|c|c}
		mesh & $\alpha_\text{ph}T^{-1}$ & $\alpha_\text{el}T^{-1}$ & $\alpha_\text{tot}T^{-1}$& $\sigma S$ & Kelvin dev. (\%)\\
		\hline
		$(30,300)$ & $-677.57$ & $-328.95$ & $-1006.5$ & $-1867.6$ & $46.1$\\
		$(45,450)$ & $-1329.6$ & $-364.75$ & $-1694.3$ & $-2053.1$ & $17.5$ \\
		$(60,600)$ & $-2134.3$ & $-488.61$ & $-2623.0$ & $-2953.0$ & $11.2$ \\
		$(90,630)$ & $-2481.0$ & $-577.14$ & $-3058.1$ & $-3374.7$ & $9.4$ \\
	\end{tabular}
	\vspace{0.5cm}\\
	\begin{tabular}{c|c|c|c|c|c}
		mesh & $\alpha_\text{ph}T^{-1}$ & $\alpha_\text{el}T^{-1}$ & $\alpha_\text{tot}T^{-1}$& $\sigma S$ & Kelvin dev. (\%)\\
		\hline
		$(30,300)$ & $-1.10$ & $-28.67$ & $-29.77$ & $-30.65$ & $2.9$\\
		$(45,450)$ & $-1.71$ & $-32.68$ & $-34.39$ & $-34.18$ & $0.6$ \\
		$(60,600)$ & $-2.47$ & $-34.15$ & $-36.62$ & $-36.27$ & $0.9$ \\
	\end{tabular}
	\caption{Breakdown of the transport coefficient $\alpha T^{-1}$ into phonon and electronic contributions and $\sigma S$. Both these coefficients are in units of A$^{-1}$m$^{-1}$K$^{-1}$. The last column gives the percentage deviation from the Kelvin-Onsager relation. Top (bottom) table section gives results for $50$ ($300$) K and $10^{18}$ cm$^{-3}$ carrier concentration.}
	\label{tab:kelvin}
\end{table}

 Table \ref{tab:kelvin} top and bottom show the contributions to $\alpha T^{-1}$ from phonons, electrons and their sum compared to $\sigma S$ as a function of the $\mb{q}/\mb{k}$ grid densities at a carrier density of $10^{18}$ cm$^{-3}$ at $50 $ K and at $300 $ K. The last column shows the deviation from the Kelvin-Onsager relation, Eq. \ref{eq:kelvin}. For $T = 300 $ K, the electronic transport coefficients, $\sigma S$ and $\alpha_{\text{el}}T^{-1}$ show good convergence with increasing grid density, differing only by around $5 \%$ going from a (45,450) grid to a (60,600) grid. The Kelvin-Onsager relation is well-satisfied for all grid densities, and the phonon contribution, $\alpha_{\text{ph}}T^{-1}$ is quite small. In contrast, at $50 $ K, $\alpha_{\text{ph}}T^{-1}$ gives the dominant contribution to $\alpha_{\text{tot}}T^{-1}$ and convergence of the transport coefficients is more challenging to achieve.  Nevertheless, we find that with increasing mesh density, this error in satisfying the Kelvin-Onsager relations diminishes. The error in $\alpha_{\text{ph}}T^{-1}$ is mainly coming from the lack of resolution of the very low energy phonons which contribute strongly at low temperatures. We will discuss this issue further below. We note that the lack of convergence of the electronic transport coefficients, $\alpha_{\text{el}}T^{-1}$ and $\sigma S$, with increasing grid density is also driven by the poor resolution of the phonon grid.  When phonons are constrained to be in equilibrium, $\alpha_{\text{el}}T^{-1}$ and $\sigma S$ show good convergence at $50 $ K differing only by about $0.5\%$ in going from a (45,450) grid to a (60,600) grid.

\begin{figure}
	\centering
	\subfloat{
		\centering
		\includegraphics[scale=0.45]{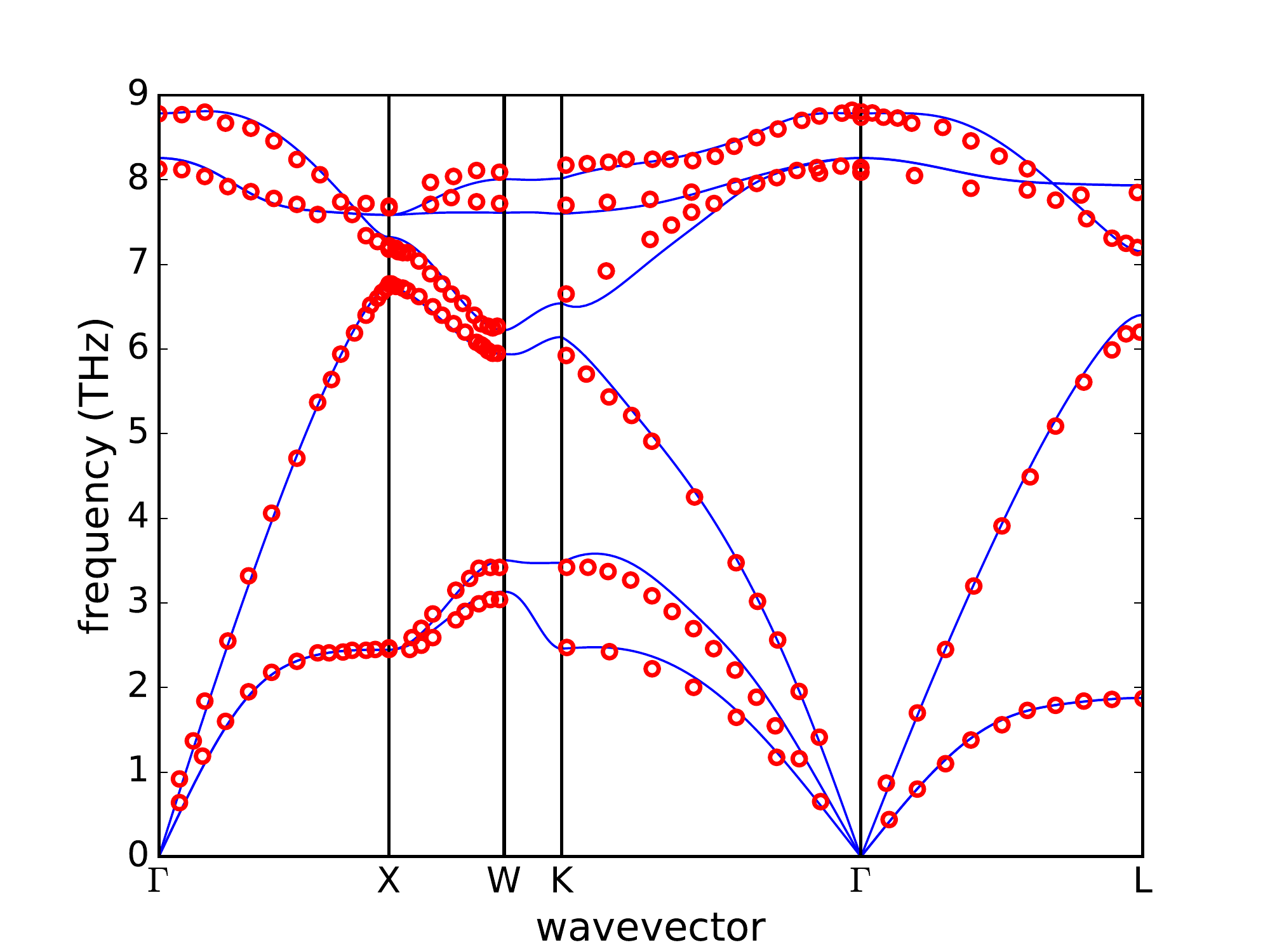}
	}	
	\caption{Phonon dispersions of GaAs. Experimental data extracted from Ref. \cite{strauch1990phonon} are plotted in red circles.}
	\label{fig:gaasdisp}
\end{figure}

\begin{figure}
	\centering
	\subfloat{
		\centering
		\includegraphics[scale=0.45]{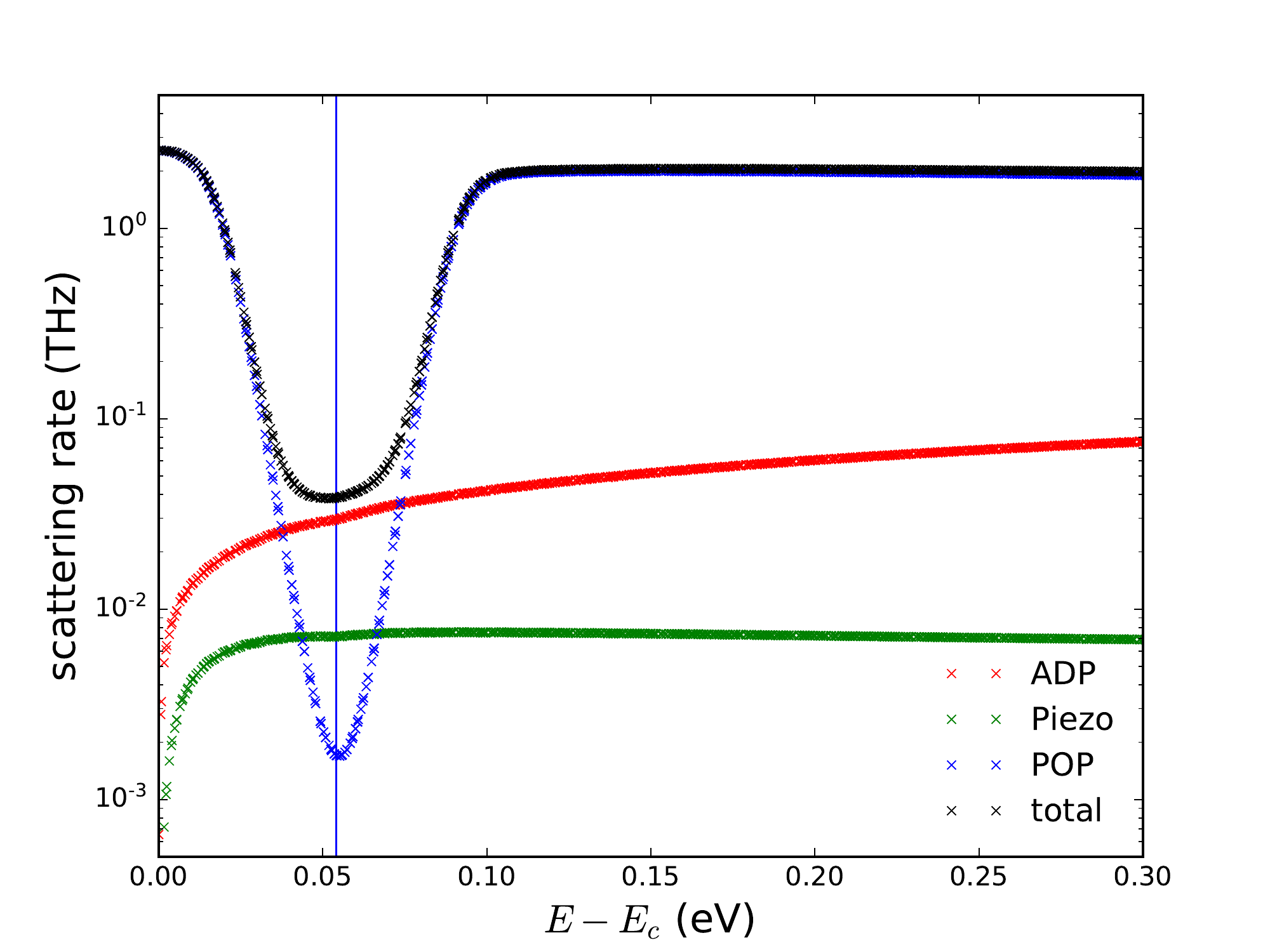}
	}
	\\
	\subfloat{
		\centering
		\includegraphics[scale=0.45]{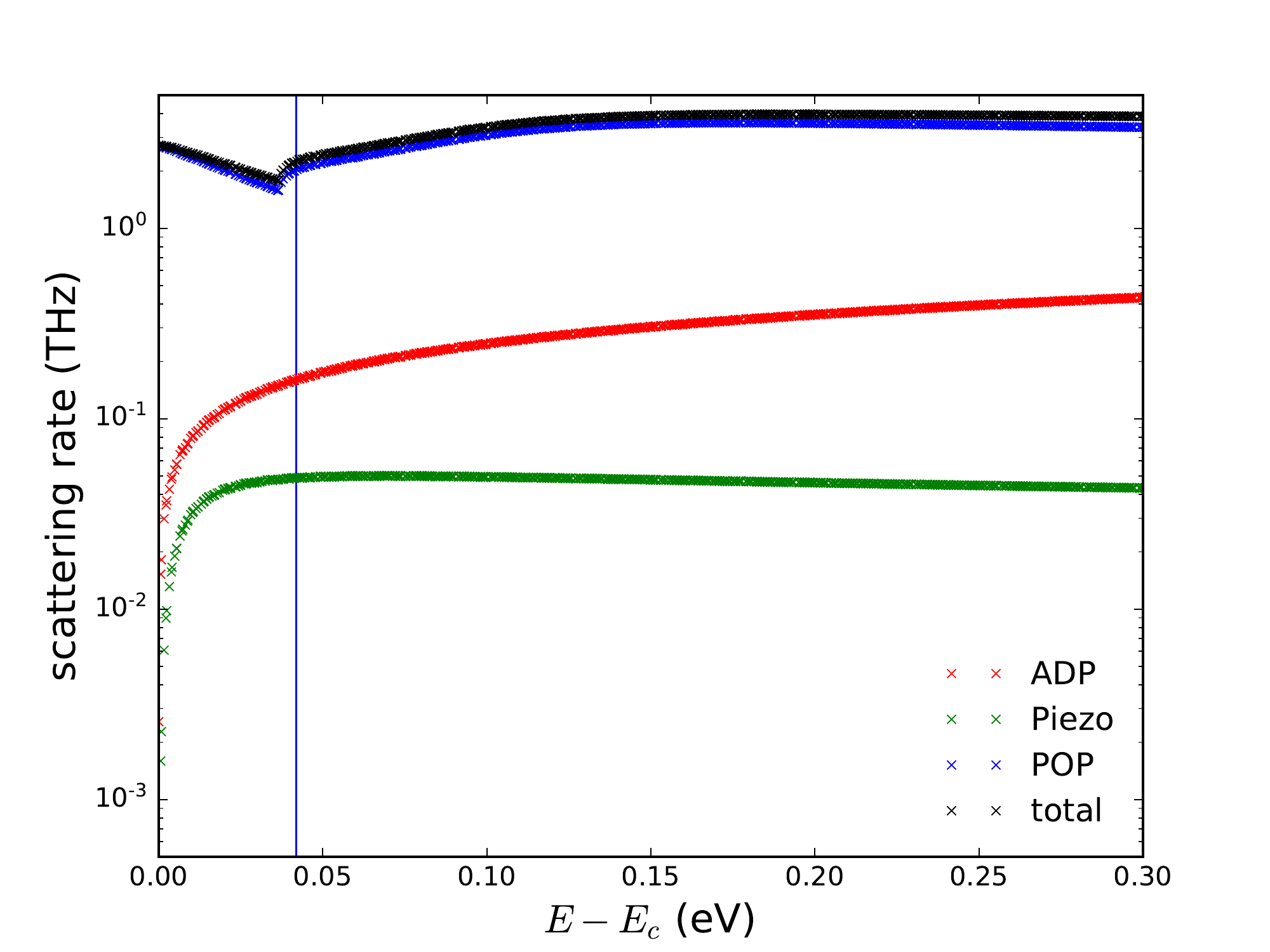}
	}	
	\caption{Total e-ph RTA scattering rates and the breakdown into various channels for $10^{18}$ cm$^{-3}$ carrier concentration. The top (bottom) panel is for $50$ $(300)$ K. The vertical line denotes the chemical potential.}
	\label{fig:gaaseph}
\end{figure}

\section{Results and discussion}
Using the local density approximation (LDA) with the Bachelet, Hamann and Schluter (BHS) pseudopotential \cite{bachelet1982pseudo}, we find the relaxed GaAs lattice constant to be $5.55$ \AA, about $1.7\%$ below the measured value of $5.65$ \AA. Such a reduction is typical of LDA calculations, which are known to overbind. The calculated phonon dispersions are shown in Fig. \ref{fig:gaasdisp} along with experimental data from Ref. \cite{strauch1990phonon}. Excellent agreement between calculation and experiment is seen.

Fig. \ref{fig:gaaseph} shows the e-ph RTA scattering rates calculated from Eqs. \ref{eq:erta}, \ref{eq:er}, and \ref{eq:ex} for the various interaction channels at $50$ K and $300$ K for a carrier concentration of $10^{18}$ cm$^{-3}$. We note that scattering of carriers by ionized impurities is not considered in the present work.  Because the $50$ K Fermi function is quite sharp near the chemical potential, an electron occupying a state near it will not be likely to jump to an energy lower by an amount equal to the $36$ meV LO phonon energy in GaAs. Similarly, approaching the chemical potential from below, the occupation number decreases sharply, causing a decrease in the optic phonon absorption. Because of this suppression of optic phonon absorption and emission, the POP rates show a strong dip near the chemical potential. This effect is not strong for small energy acoustic phonon scattering, as energy levels infinitesimally close to the chemical potential have similar occupation numbers, which is why the e-acoustic phonon rates dominate in that region. Thus, at $50$ K, the long lived electrons near the chemical potential predominantly exchange momenta with the low energy acoustic phonons. As the Fermi function is smeared out at $300$ K, the strong dipping of the POP scattering rates does not happen. As such, at such high temperatures, electrons predominantly exchange momenta with the LO phonons.

\begin{figure}
	\centering
	\subfloat{
		\centering
		\includegraphics[scale=0.45]{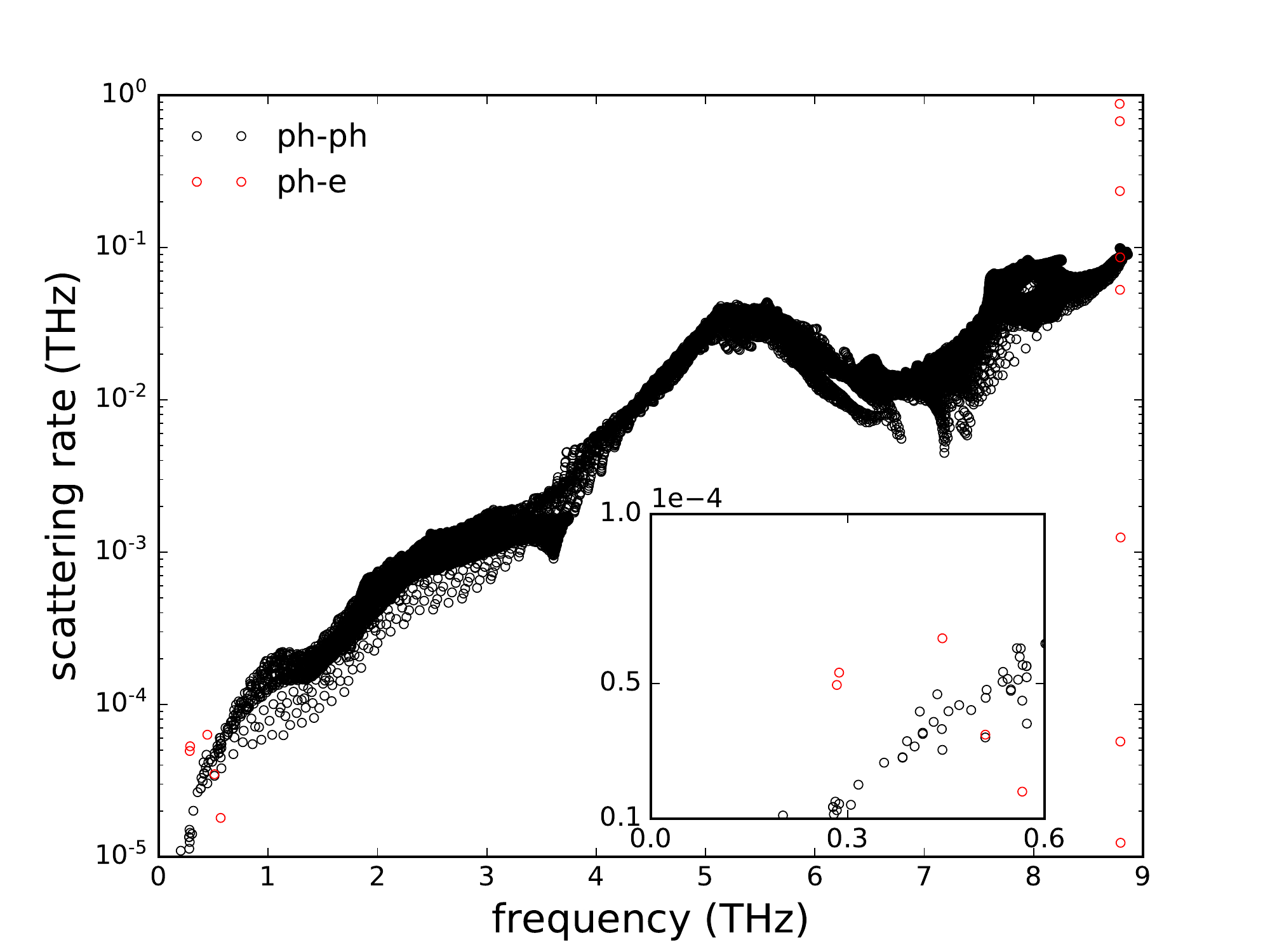}
	}
	\\
	\subfloat{
		\centering
		\includegraphics[scale=0.45]{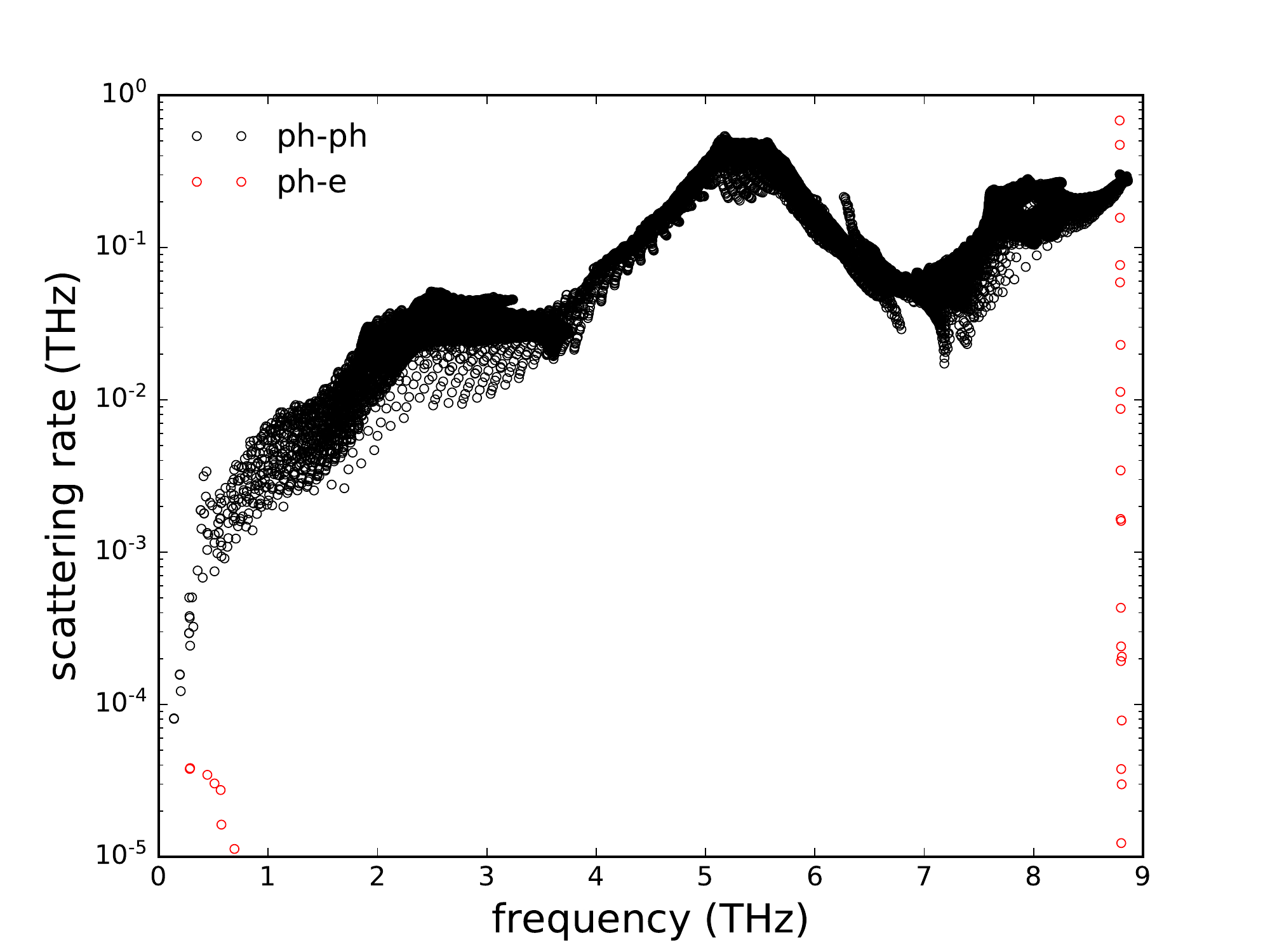}
	}	
	\caption{Ph-ph and ph-e scattering rates for $10^{18}$ cm$^{-3}$ carrier concentration. The top (bottom) panel is for $50$ $(300)$ K. Inset in top panel shows low frequency ph-el and ph-ph scattering rates.}
	\label{fig:gaasphe}
\end{figure}

Next, we look at the ph-ph and ph-e scattering rates for a carrier concentration of $10^{18}$ cm$^{-3}$ at $50$ K and $300$ K in Fig. \ref{fig:gaasphe}. We first note that as the temperature is lowered, ph-e scattering rates get stronger. This is again related to the sharpness of the Fermi function. At low temperatures, an electron near the chemical potential has nearly empty higher energy states available. It can then easily make this transition by absorbing acoustic or optic phonons. Anharmonic phonon scattering rates, in contrast, decrease with decreasing temperature as there are fewer phonons available to scatter with. Now, comparing to the e-ph scattering rates in Fig. \ref{fig:gaaseph}, we see that the electronic momenta that were pumped predominantly into the very small energy acoustic phonons at $50$ K ($< 1$ THz), can flow back into the electronic system as ph-e scattering rates are higher than ph-ph scattering rates for these phonons. In contrast, at $300$ K, ph-e rates are smaller than ph-ph rates for acoustic phonons and only the LO phonons remain drag active. Thus, we expect a stronger drag effect at low temperatures.

\begin{figure}
	\centering
	\subfloat{
		\centering
		\includegraphics[scale=0.45]{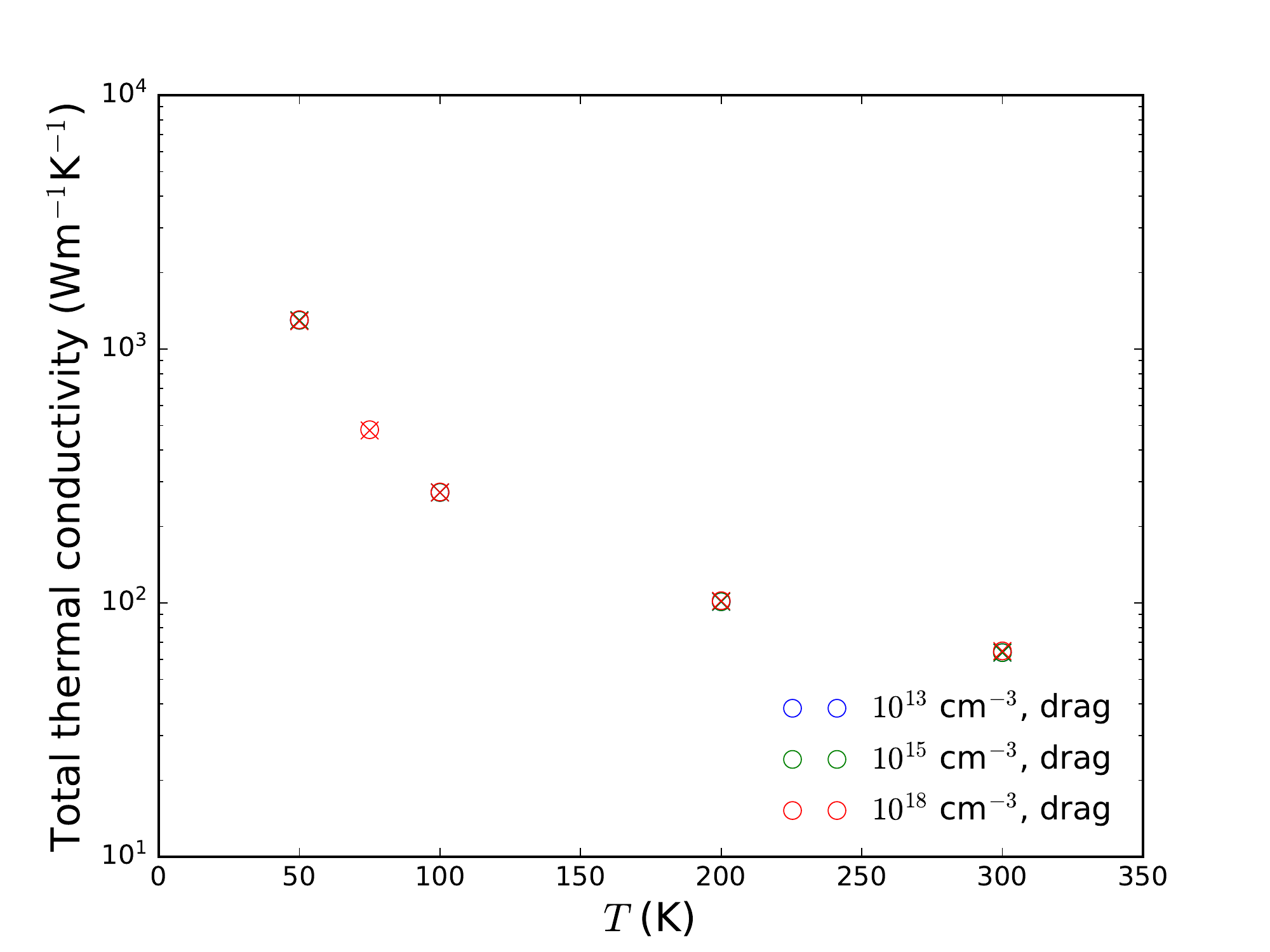}
	}	
	\caption{Temperature dependence of the total thermal conductivity including electronic and phonon contributions. Open circles (crosses) for a given color denote the drag (non-drag) total thermal conductivity.}
	\label{fig:gaaskappa}
\end{figure}

\begin{figure}
	\centering
	\subfloat{
		\centering
		\includegraphics[scale=0.45]{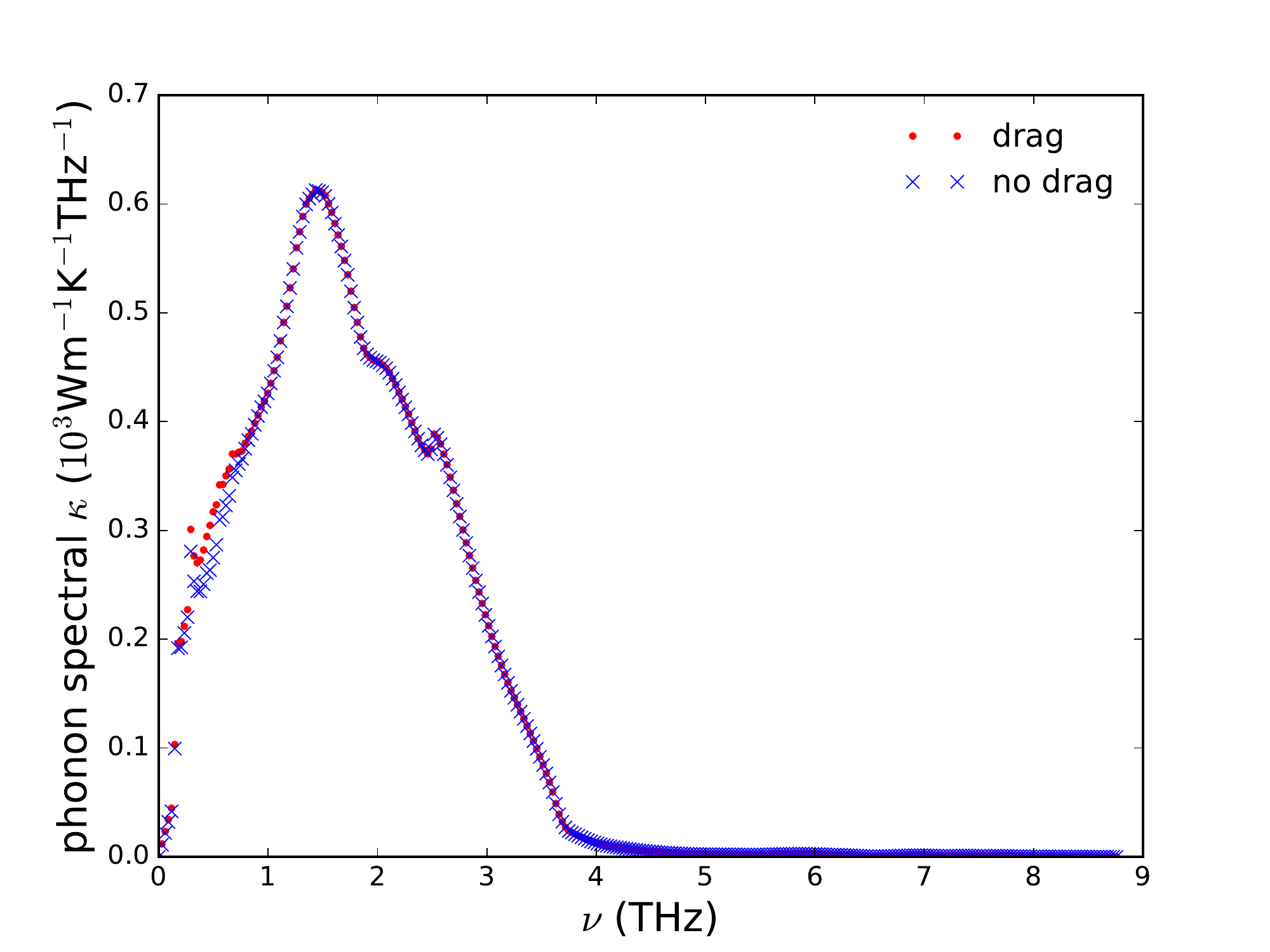}
	}	
	\caption{Spectral phonon thermal conductivity at $50$ K for a carrier concentration of $10^{18}$ cm$^{-3}$.}
	\label{fig:gaasphknu}
\end{figure}

Now we study the effect of drag on the various transport coefficients. We start by looking at the temperature dependence of the total thermal conductivity in Fig. \ref{fig:gaaskappa}. Blue, green and red circles (crosses) denote the thermal conductivity taking into account (neglecting) drag for $10^{13}$, $10^{15}$, and $10^{18}$ cm$^{-3}$ carrier concentrations. The electronic thermal conductivity is nearly two orders of magnitude smaller than the phonon contribution as was shown in Table \ref{tab:conv}. The effect of drag on the phonon thermal conductivity is weak since the spectral contribution comes from a wider range of phonon frequencies, and not just from the strongly drag active low energy acoustic and high energy optic phonons. The spectral contribution to the lattice thermal conductivity at $50$ K and $10^{18}$ cm$^{-3}$ carrier concentration is given in Fig. \ref{fig:gaasphknu}, where the red dots (blue crosses) denote the calculation with (without) drag. Only around $0.5$ THz do we see a slight increase in the thermal conductivity which can be connected to the dominant ph-e scattering around that energy scale in the top panel of Fig. \ref{fig:gaasphe}. The high energy, and low speed optic phonons contribute negligibly to the thermal conductivity. In Refs. \cite{liao2015,protik2017,wangsic,li2018fermi} it was assumed that the electron system remains in equilibrium following interaction with the phonons. Here we see the justification for making that assumption.

\begin{figure}
	\centering
	\subfloat{
		\centering
		\includegraphics[scale=0.45]{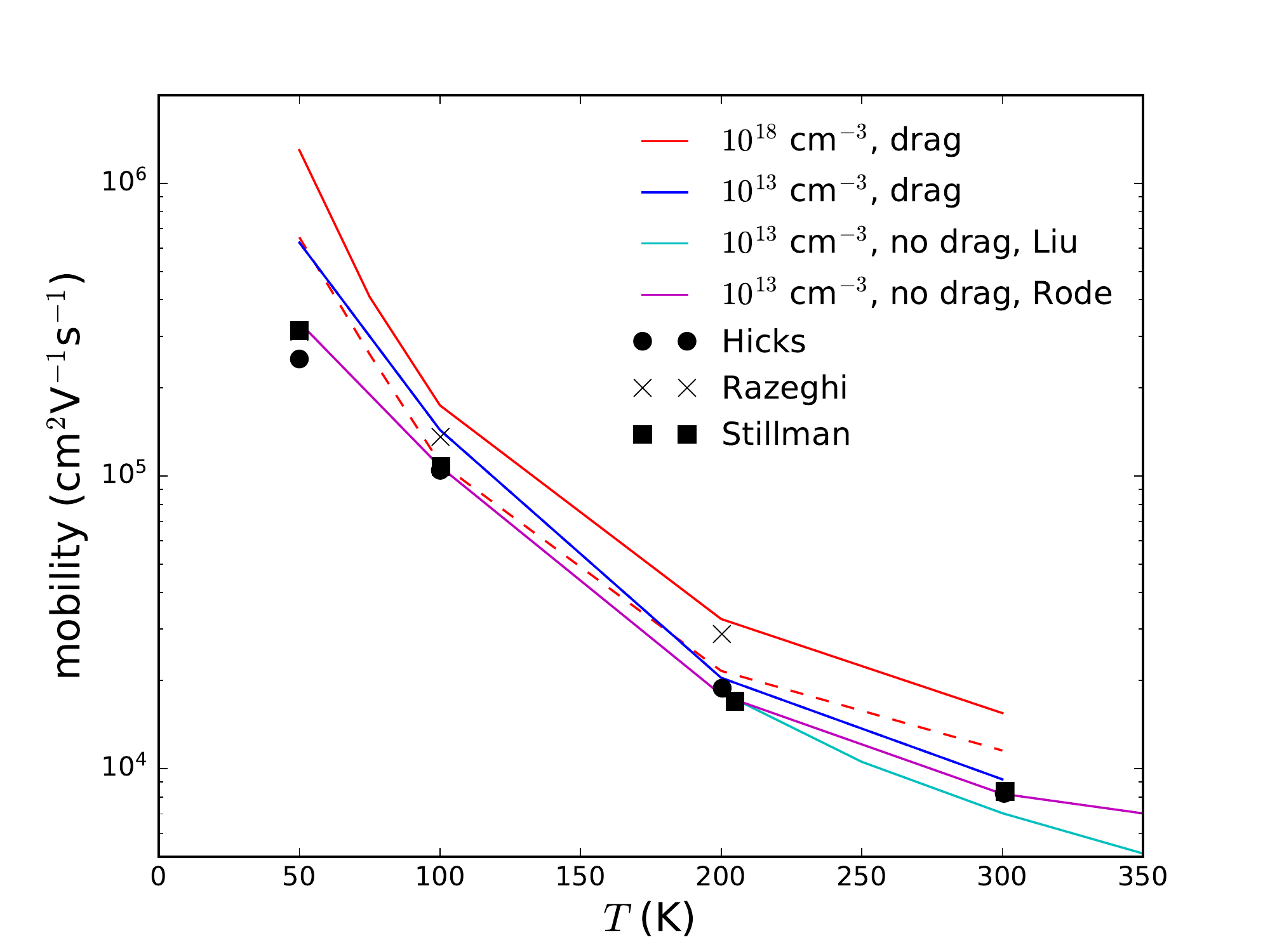}
	}	
	\caption{Temperature dependence of the carrier mobility. Red solid (dashed) line gives the mobility due to $10^{18}$ cm$^{-3}$ carrier concentration with (without) phonon drag. Other curves are for $10^{13}$ cm$^{-3}$ carrier concentration including calculated results from Refs. \cite{rode1971electron,liu2017first}, compared to experiments from Refs. \cite{hicks1969high,stillman1970hall,razeghi1989high}.}
	\label{fig:gaasmob}
\end{figure}

\begin{figure}
	\centering
	\subfloat{
		\centering
		\includegraphics[scale=0.45]{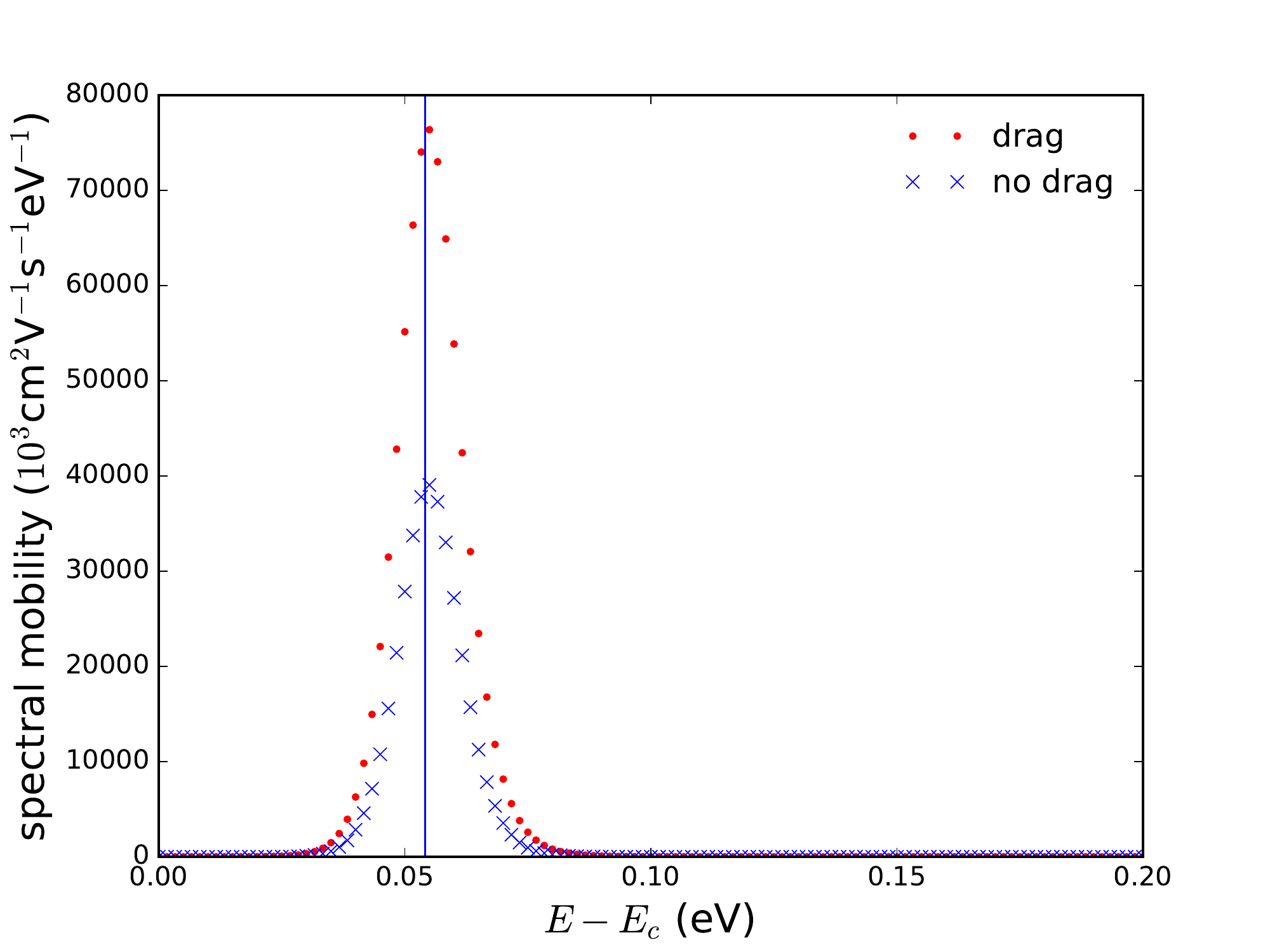}
	}	
	\caption{Spectral mobility at $50$ K for a carrier concentration of $10^{18}$ cm$^{-3}$. The vertical blue line marks the chemical potential.}
	\label{fig:gaasmobnu}
\end{figure}

The carrier mobility, on the other hand, shows a large gain due to drag for high concentrations as shown in Fig. \ref{fig:gaasmob}. The red solid (dashed) line gives the mobility due to $10^{18}$ cm$^{-3}$ carrier concentration with (without) phonon drag. The mobility for the $10^{13}$ cm$^{-3}$ carrier concentration is given by the blue line. The result for an intermediate concentration of $10^{15}$ cm$^{-3}$ gives very similar result to the $10^{13}$ cm$^{-3}$ case, and is not plotted here. For these low concentration cases the drag effect is negligible. Experimental data \cite{hicks1969high,stillman1970hall,razeghi1989high} for nearly intrinsic ($\approx 10^{13}$ cm$^{-3}$) samples are plotted in black symbols. The maroon line gives analytical model calculation results disregarding the drag effect by Rode and Knight \cite{rode1971electron}. The cyan line shows first principles DFT+BTE results from Liu \cite{liu2017first}, where drag has also been ignored. The first principles DFT method has also been employed by Zhou and Bernardi in Ref. \cite{zhou2016ab}. There the RTA has been used and drag was ignored. Both Refs. \cite{liu2017first} and \cite{zhou2016ab} obtain good agreement with the measured data at low carrier concentration ($10^{13}$ cm$^{-3}$), but there is significant disagreement between these two fully \emph{ab initio} calculations: Liu et. al. found that the full solution of the BTE is required to get good agreement with experiments, while Zhou and Bernardi found that the RTA is sufficient to match the experimental values. We note that there exist important methodological differences between these two studies. First, Liu et. al. employed the GW method to ensure accurate calculation of the conduction band mass, while Zhou and Bernardi used the standard LDA functional and do not discuss effective mass in their work. Secondly, Liu et. al. use the analytic tetrahedron method \cite{lambin1984computation}, while Zhou and Bernardi use the Lorenzian broadening method for approximating the energy conserving delta functions. Note that our calculated mobilities at $10^{13}$ cm$^{-3}$, which include drag, (blue curve) are very close to the \emph{ab initio} results in Refs. \cite{liu2017first} and \cite{zhou2016ab}. This is to be expected since for this low carrier density, drag effects are negligible, the parabolic band approximation for GaAs is quite accurate, and the parameters defining the electron-phonon scattering have already been adjusted to fit the measured data \cite{rode1970electron}.

At low concentrations, the ph-e scattering rates are weaker than ph-ph scattering rates and as such the drag effect on mobility is weak. This is why ignoring the drag effect was justified in Refs. \cite{rode1971electron,liu2017first,zhou2016ab}. For higher carrier densities, such as the $10^{18}$ cm$^{-3}$ case considered here, as temperature is lowered, the strong reduction of the POP scattering rates near the chemical potential causes the increase in the mobility. In the absence of impurities, higher carrier concentration gives stronger screening, weakening the e-ph scattering rates and increasing the mobility. Agreement of the low concentration data with our $10^{13}$ and $10^{15}$ cm$^{-3}$ results are very good throughout the considered temperature range. Near $50$ K, impurity scattering is already important, which is why our calculated results are higher than the experimental results and Rode's analytic calculation that includes impurity scattering. For the $10^{18}$ cm$^{-3}$ concentration case, we find that the phonon-drag effect on mobility is strong for all temperatures - the mobility drag gain is $99\%$, $57\%$, $44\%$, and $34\%$ at $50$, $100$, $200$ and $300$ K, respectively. The drag gain mostly comes from near the chemical potential, as shown in the spectral mobility plot in Fig. \ref{fig:gaasmobnu}, which is a direct consequence of the exchange of electronic momenta with the small energy acoustic phonons for the low temperature case and large energy optic phonons for the high temperature case. As explained earlier, the increase in the drag effect with the lowering of temperature is the expected behavior. However, this effect on mobility may not be observed at low temperatures since impurity scattering will be a dominant electron scattering channel. If, however, carriers can be introduced into a sample without introducing impurities, such as by modulation doping \cite{stormer1979two}, we predict that the low temperature mobility should show a noticeable enhancement due to phonon drag.

\begin{figure}
	\centering
	\subfloat{
		\centering
		\includegraphics[scale=0.45]{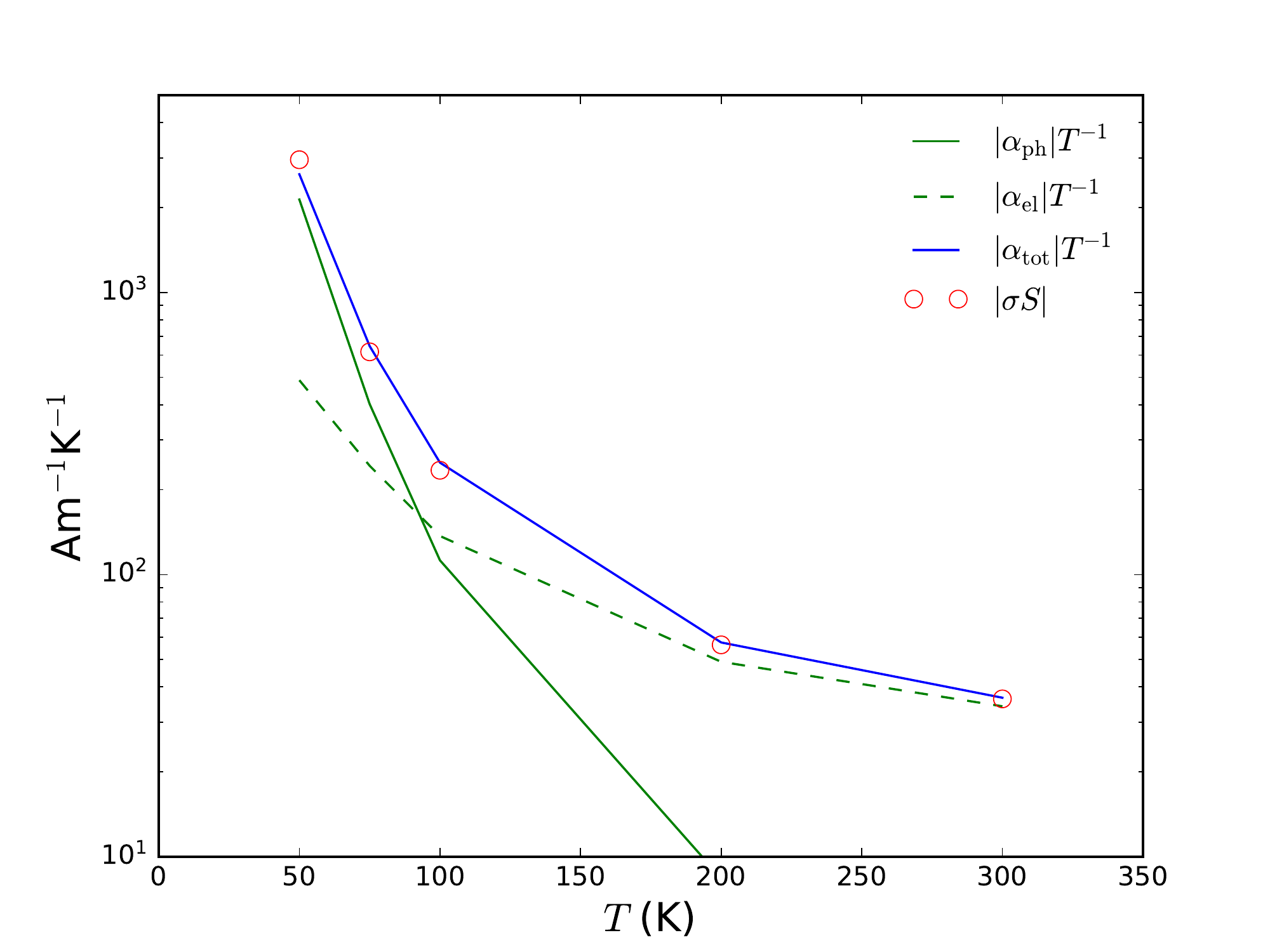}
	}	
	\caption{Temperature dependence of $\sigma S$ and the phonon and electronic contributions to $\alpha_{\text{tot}}T^{-1}$. The plotted quantities are for $10^{18}$ cm$^{-3}$ carrier concentration.}
	\label{fig:kelvin}
\end{figure}

\begin{figure}
	\centering
	\subfloat{
		\centering
		\includegraphics[scale=0.45]{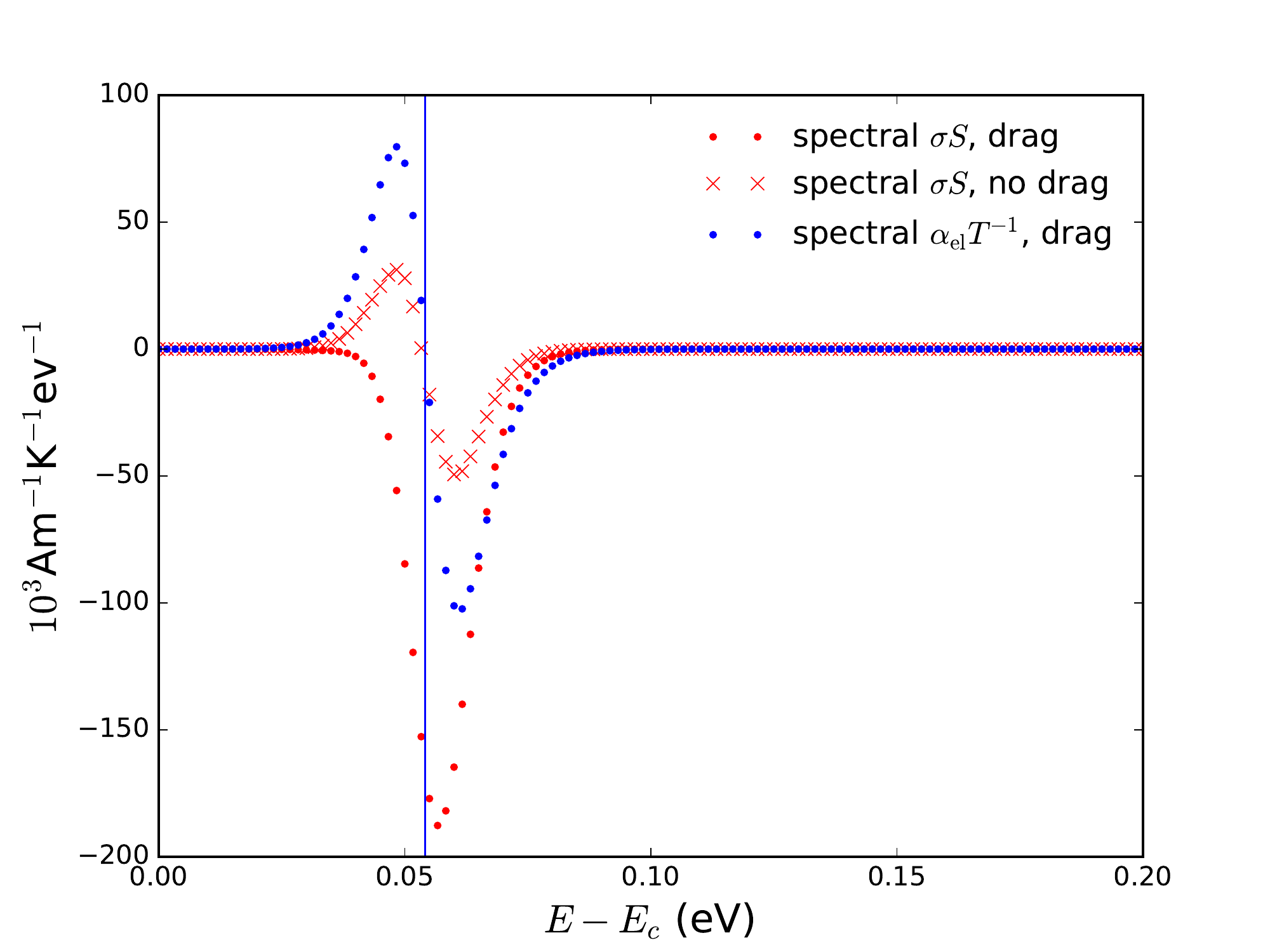}
	}
	\\
	\subfloat{
		\centering
		\includegraphics[scale=0.45]{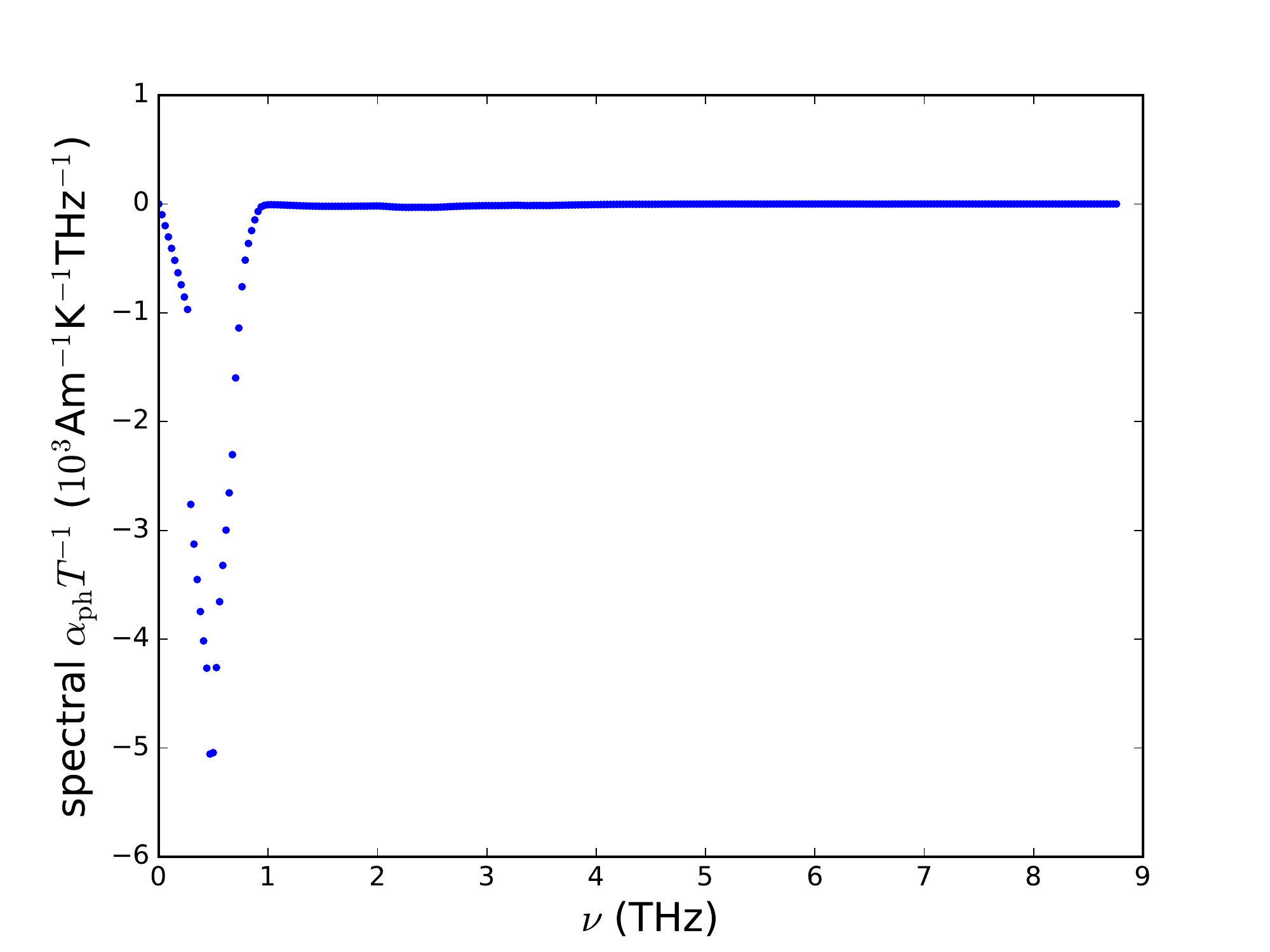}
	}	
	\caption{Top panel: spectral $\sigma S$ and electronic contribution to $\alpha_{\text{tot}}T^{-1}$. Vertical line shows the location of the chemical potential. Bottom panel: spectral phonon contribution to $\alpha_{\text{tot}}T^{-1}$. Both plots are for $50$ K and  $10^{18}$ cm$^{-3}$ carrier concentration with drag. Upper plot also includes spectral $\sigma S$ without drag.}
	\label{fig:sigmaSnu}
\end{figure}

Before discussing the effect of drag on the Seebeck coefficient, we look at the transport quantities $\sigma S$ and $\alpha_{\text{tot}} T^{-1}$. Ideally, these two quantities are equal, which is the statement of the Kelvin-Onsager relation. As shown in Table \ref{tab:kelvin}, we find that the deviation from this relation at $50$ K diminishes with increasing mesh density. In Fig. \ref{fig:kelvin} we show the temperature dependence of these quantities. We note that at room temperature down to $200$ K, the phonon contribution to $\alpha_{\text{tot}} T^{-1}$ is quite small reflecting the dominance of phonon-phonon decay over phonon-electron scattering. However, as we continue to lower the temperature, the direct phonon contribution to $\alpha_{\text{tot}} T^{-1}$ starts to dominate over the direct electron contribution. In this regime, low energy phonons are strongly moved out of equilibrium due to the dominance of electron-acoustic phonon scattering near the chemical potential and the dominance of phonon-electron scattering compared to phonon-phonon scattering for the low frequency phonons. For the $600\times600\times600/60\times60\times60$ $\mb{k}/\mb{q}$ grids, the numerical agreement with the Kelvin-Onsager relation is $11.18\%$, $4.57\%$, $6.04\%$, $1.92\%$ and $0.94\%$ at $50$, $75$, $100$, $200$ and $300$ K, respectively. While $\sigma S$ and $\alpha_{\text{el}}T^{-1}$ are quantities calculated by summing over the electronic wavevectors, the evaluation of $\alpha_{\text{ph}}T^{-1}$ requires summing over the phonon wavevectors. As a large spectral contribution of $\alpha_{\text{ph}}T^{-1}$ comes from the low energy acoustic phonons, as can be seen in the bottom panel of Fig. \ref{fig:sigmaSnu}, a finer $\qq$-mesh is required for better agreement with the Kelvin-Onsager relation. In our current scheme, a $\qq$-mesh much finer than $60\times 60\times 60$ is computationally prohibitive, and we relegate the circumvention to this numerical problem for the future. In contrast, the spectral contribution to $\alpha_{\text{el}}T^{-1}$ and $\sigma S$ comes mostly from near the chemical potential, as shown in the top panel of Fig. \ref{fig:sigmaSnu}. This energy range is well resolved by the $600 \times 600 \times 600$ $\kk$-mesh used.

It is interesting to  note the dramatically different form of the spectral contribution to $\sigma S$ in Fig. \ref{fig:sigmaSnu}, obtained by solving the coupled BTEs (i.e. including phonon drag), compared with that obtained when phonons are constrained to stay in equilibrium. In the latter case, the spectral $\sigma S$ is nearly anti-symmetric about the chemical potential, $\mu$, giving only small $\sigma S$. Thus, the contributions to $\sigma S$ in Fig. \ref{fig:sigmaSnu} are coming almost exclusively from the phonon drag effect, and they exhibit a roughly symmetric form about $\mu$.

\begin{figure}
	\centering
	\subfloat{
		\centering
		\includegraphics[scale=0.45]{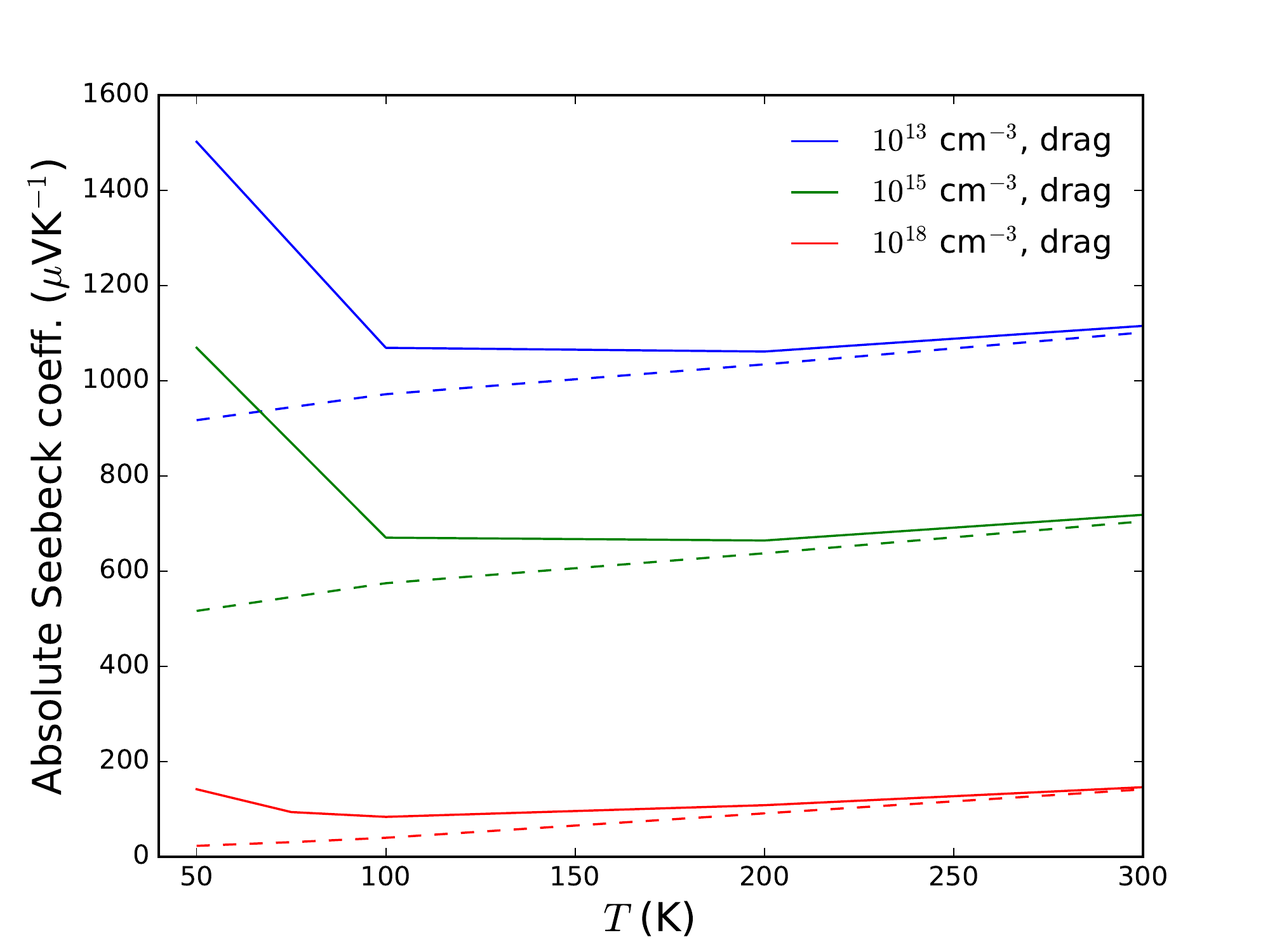}
	}	
	\caption{Temperature dependence of the absolute value of the carrier Seebeck coefficient, $|S|$ for various carrier concentrations.}
	\label{fig:gaasseebeck1}
\end{figure}

\begin{figure}
	\centering
	\subfloat{
		\centering
		\includegraphics[scale=0.45]{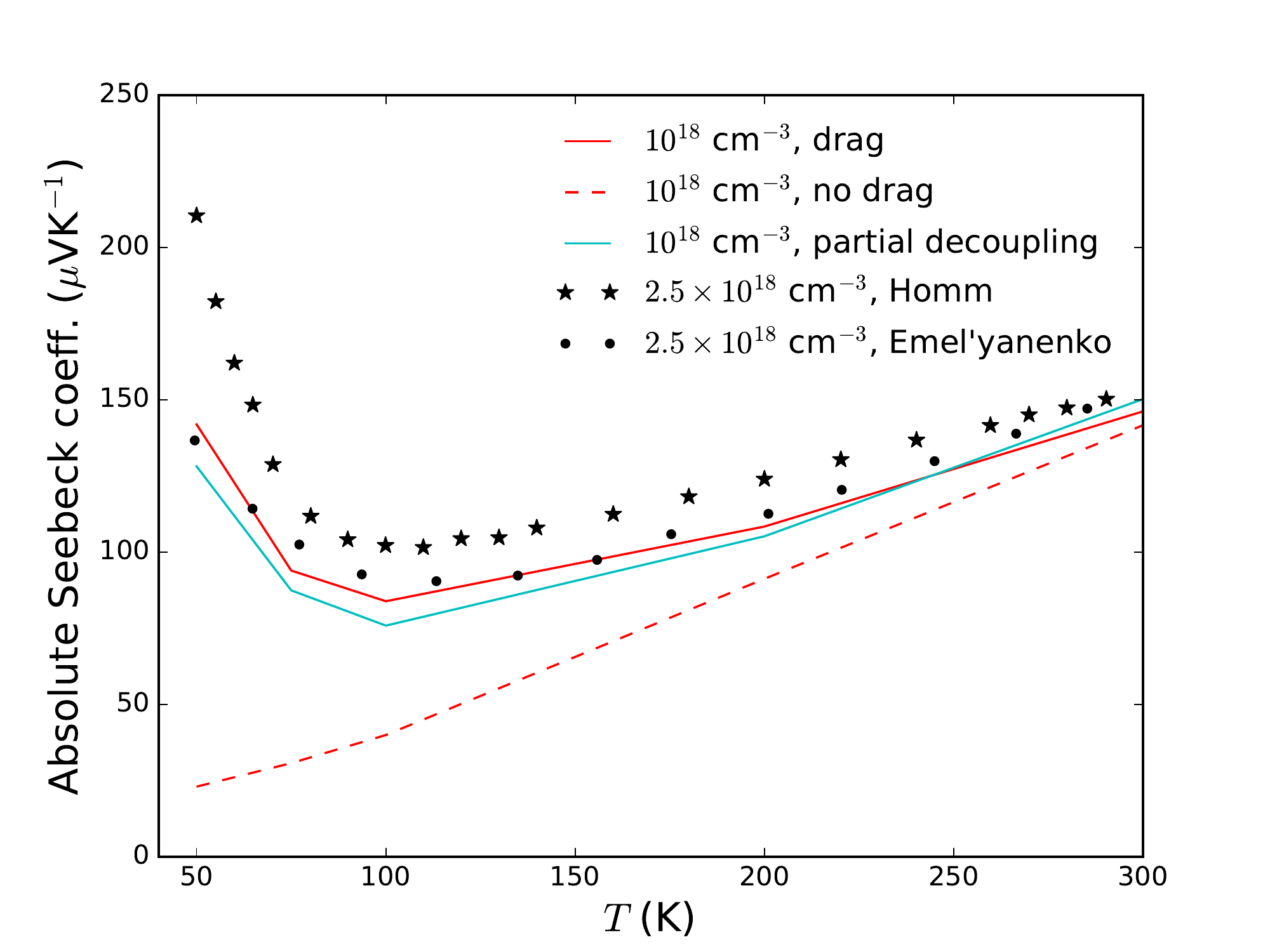}
	}	
	\caption{Temperature dependence of the absolute value of the carrier Seebeck coefficient, $|S|$ for high carrier concentration. Inclusion of phonon drag results in the minimum near $100$ K, which is also seen in the experimental data \cite{homm2008,gaascollection}.}
	\label{fig:gaasseebeck2}
\end{figure}

Lastly, we consider the Seebeck coeffcient in Fig. \ref{fig:gaasseebeck1}.  Blue, green and red solid (dashed) lines denote the modulus of the Seebeck coefficient taking into account (neglecting) drag for $10^{13}$, $10^{15}$, and $10^{18}$ cm$^{-3}$ carrier concentration. All drag curves show a turn-over of $|S|$ near $100$ K. We compare the high concentration case with experimenal data in Fig. \ref{fig:gaasseebeck2}. Experimental data for $2.5\times 10^{18}$ cm$^{-3}$ carrier concentration \cite{homm2008,gaascollection} are plotted in black symbols. We find that drag theory faithfully reproduces the upturn of $|S|$ near $100$ K, whereas the non-drag theory spectacularly fails to capture this effect. It is noteworthy that despite the use of simple models for the e-ph matrix elements, the calculated drag theory $|S|$ is in qualitative agreement with the experimental data. This effect is also captured well by the partially decoupled BTE solution, given by the cyan line. In this latter approach, applied to silicon in Refs. \cite{mahan2014seebeck,fiorentini2016thermo}, the phonon drag term in the electron BTE, Eq. \ref{eq:TebteD}, is set to the RTA value, Eq. \ref{eq:Tpbte0}, for all iterations. As such, in this approximation one does not solve the coupled set of electron-phonon BTEs, but treats the drag effect approximately. We note that use of an RTA could lead to the Kelvin-Onsager relation being badly violated, depending on the choice of the relaxation time. This method predicts the turn-over of $|S|$ very well. The success of the partial decoupling method in this case is due to the fact that in GaAs, (i) the RTA is a good approximation for the low frequency out-of-equilibrium phonons that contribute strongly to drag, and (ii) the electron drag effect is weak. The phonon-drag effect on $|S|$ can  be understood in the following way. The fixed temperature gradient $\mb{\nabla} T$ generates an initial electronic heat current down the gradient. On the other hand, since the circuit is open, the net accumulation of the electrons at the cold end generates an electric field opposite to the temperature gradient. So, in steady state, a constant electric field, $\mb{E}$, develops. The Seebeck coefficient is defined through $\mb{E} = S\mb{\nabla} T$. $|S|$ is large if the final induced field is large, which can happen if more electrons can travel from the hot to the cold side working against the induced field that is developing during the initial transient period. If phonons are allowed to go out of equilibrium, which is important when electrons and phonons can transfer quasi-momentum faster than it can be dissipated by anharmonic phonon decay i.e. at sufficiently low T, there is a flow of phonons down the temperature gradient. This $\mb{\nabla}T$-downstream phonon ``wind", conveyed through the electron-phonon interaction, helps more electrons to overcome the growing electric field. Thus, the enhanced phonon-drag at low temperatures leads to the strong enhancement of $|S|$. In the absence of phonon-drag, $|S|$ decreases monotonically with decreasing temperature, contrary to what is observed experimentally. This happens because of the increasing cancellation of contributions to the electric current from carriers above and below the chemical potential as temperature is lowered. 

\section{Summary and outlook}

In this work, we constructed a full numerical solution of the coupled electron-phonon Boltzmann equations for charge and heat transport and considered n-doped GaAs as an example case. We combined \emph{ab initio} phonon dispersions and phonon-phonon matrix elements with model electronic band and electron-phonon matrix elements to assess the effects of the mutual drag that electrons and phonons exert on each other on various transport coefficients. The main findings are: 1) Phonon thermal conductivity is unaffected by the drag effect. This happens because the spectral contribution to the phonon thermal conductivity comes from a large frequency range, and the drag effect only provides a gain in the low frequency regime. To significantly affect the thermal conductivity of a moderately doped semiconductor, the phonon-electron scattering rates would need to exceed phonon-phonon scattering rates over a wider range of frequencies. 2) Provided that the electron-impurity scattering can be made weak, the carrier mobility is strongly enhanced by the drag effect at high carrier concentrations ($\approx 10^{18}$ cm$^{-3}$) and at low temperatures, which is a consequence of strong momentum exchange between electrons near the chemical potential and low energy phonons. And, 3) There is a strong enhancement of the absolute value of the Seebeck coefficient at low temperature leading to good agreement between the drag theory and experiment. The non-drag transport theory increasingly fails with decreasing temperatures with catastrophic failure at low T.

This analytic model+ab initio hybrid calculation also provides us with some insight about the various levels of rigor needed when approaching a transport problem. For example, by looking at the e-ph, ph-e and ph-ph scattering rates at a given temperature, we can tell whether the non-drag theory will give us a good result for a transport coefficient; if not, we need to use the computationally expensive drag theory. Moreover, our hybrid calculation demonstrates that strong drag effects can be captured at an \emph{ab initio} level, motivating development of a completely parameters-free implementation of the code, which could prove useful for materials research.

We note that analytical models for electron-phonon scattering and conduction band structure similar to the one used here for GaAs could be easily accommodated for other materials.  The only parameters that enter the model are the acoustic deformation potential, the piezoelectric constant, the static and high frequency dielectric constants and the conduction band effective mass. These parameters have been tabulated for other direct-gap semiconductors (see e.g. Ref. \cite{rode1970electron} for GaSb, InP, InAs, InSb) and could be included without any change to the presented computational framework. Even indirect gap semiconductors have been investigated using similar analytical models(see e.g. Ref. \cite{rode1972electron}). Treatment of the multiple anisotropic conduction band valleys in Si and Ge would introduce additional computational expense but is in principle possible within the developed framework.

In order to go to temperatures lower than $50$ K, we need to address the numerical issue of insufficiently resolving the low energy acoustic phonons which leads to the worsening of the agreement with the Kelvin-Onsager relation. One possible remedy to this problem could be to employ an additional ultrafine phonon wavevector mesh near the $\Gamma$-point. We also ignored the effect of impurity scattering in our gallium arsenide study as our goal was to understand the mutual drag of interacting electrons and phonons. Going forward, a rigorous treatment of electron-charged defect scattering within the first principles diagrammatic approach can be done. One such study was recently carrier out in Ref. \cite{lu2019efficient} using the Born approximation for neutral defects. The treatment of charged defects within a supercell approach has some additional complications such as the long-range interaction between periodic images of the defect and compensating opposite jellium charge added by DFT for charge neutrality. There are various methods that have been developed to address these problems \cite{freysoldt2014first,lany2008assessment}, and it is important to implement these in order to capture the effect of charged defect scattering accurately. We also ignored electron-electron scattering in our study. We discussed in the body of the text the difficulty in treating electron-electron scattering within an RPA+BTE framework. However, for high carrier concentrations, dynamic screening and plasmonic effects cannot be a priori ignored. The open question is: How can we accurately treat electron-electron interactions within a computationally feasible transport framework? We wish to investigate this question in the future.

\section*{Acknowledgements}
This work was supported by the Office of Naval Research under MURI grant No. N00014-16-1-2436. We acknowledge the Boston College Linux clusters for computational resources and support. NHP acknowledges helpful discussion with Dr. Navaneetha Krishnan Ravichandran and Dr. Chunhua Li.

\bibliography{References}
\end{document}